\documentclass[twocolumn,
               preprintnumbers,
               superscriptaddress,
               numbers,
               floatfix,
               nofootinbib,
               showpacs,
               colorlinks,
               linkcolor=blue,   
               citecolor=blue]{revtex4-2}

\usepackage{graphicx}
\graphicspath{{Images/}}
\usepackage{physics}   
\usepackage{adjustbox} 
\usepackage{wrapfig}
\usepackage{float}
\usepackage[utf8x]{inputenc}
\usepackage{amsmath}
\usepackage{amssymb}
\usepackage{orcidlink}
\usepackage{amsthm}
\usepackage{mathtools}
\usepackage{color}
\usepackage{multirow}
\usepackage{tabularx}
\usepackage{comment}
\usepackage{soul}
\usepackage{siunitx}
\usepackage{booktabs}
\DeclareSIUnit\bar{bar}

\usepackage{slashbox}
\usepackage{aas_macros}

\begin{document}
\title{Niebla: an open-source code for modeling the extragalactic background light}

\author{Sara Porras-Bedmar\orcidlink{0009-0001-3512-2628}}
\email{sara.porras.bedmar@uni-hamburg.de}
\affiliation{Institute for Experimental Physics, University of Hamburg, Luruper Chaussee 149, D-22761 Hamburg, Germany}

\author{Manuel Meyer\orcidlink{0000-0002-0738-7581}}
\email{mey@sdu.dk}
\affiliation{CP3-Origins, University of Southern Denmark, Campusvej 55, Odense M, DK-5230, Denmark}

\date{\today}

\begin{abstract}
   The flux of extragalactic gamma rays is attenuated through interactions with optical and infrared photons of the extragalactic background light (EBL). The EBL is an isotropic, diffuse photon field that is difficult to measure directly at these wavelengths due to strong foreground emission.
   We present \textsc{niebla}, the first open-source code to compute the EBL from optical to far-infrared wavelengths using a phenomenological approach that accepts fully customizable inputs. This software enables a detailed modeling of the influence of EBL optical depth on gamma-ray observations and facilitates the distinction between different dust reemission models.
   The code models the optical background primarily from stellar emission, by evolving the spectrum of a single stellar population as a function of redshift, considering mean metallicity evolution and star formation rate density. Additional sources to the EBL can be provided by the user. The code already includes optional contributions from, e.g., stripped stars, intra-halo light, or the decay of axion dark matter. The optical emissivity is then absorbed by interstellar dust and reemitted in the infrared regime. We provide multiple prescriptions to model this process, using spectral dust templates or a combination of blackbodies.
   We provide three EBL models calculated with different dust reemission prescriptions, which have been fitted to various observational data sets. These include EBL intensities, galactic emissivities, stellar formation rate densities, and mean metallicity of the Universe.
   In addition, we showcase the versatility of our model through a simulated observation of the blazar Markarian~501 in a high-flux state with the Large High Altitude Air Shower Observatory array. We find that the simulated VHE spectrum is highly sensitive to the photon density of the EBL at infrared wavelengths. Our model will therefore allow the community to distinguish between different dust reemission models and constrain EBL parameters with future observations.
\end{abstract}

\maketitle


\section{Introduction}
\label{sec:Intro}

The cosmic optical and infrared backgrounds (COB and CIB, respectively) constitute homogeneous and isotropic radiation fields permeating the Universe. The combination of these two backgrounds is sometimes referred as the extragalactic background light \citep[EBL; see, e.g.,][]{2001ARA&A..39..249H}, which is itself a major constituent of the cosmic photon background \citep[see, e.g.,][]{hill_spectrum_2018}. 
\footnote{Throughout this work, we adopt the definition of the EBL as encompassing the COB and the CIB.}
The EBL originates from the integrated emission of several astrophysical sources and depends sensitively on their cosmic evolution and properties \citep[see, e.g.;][]{kneiske_implications_2002,2004A&A...413..807K,haardt_radiative_2012,driver_measurements_2016,finke_modeling_2022}.

Measuring the EBL intensity directly is challenging due to the significant contamination from foreground sources.
Many of these foregrounds originate within the Solar system, such as Earth's atmosphere or Zodiacal Light, i.e., sunlight absorbed and reemitted by interplanetary dust \citep{1998HauserCib}. Other foregrounds are Galactic in origin, such as interstellar gas. Various methodologies have been developed to subtract these foreground contributions. However, the difficulty of accurately removing residual foregrounds means that, in practice, most current direct measurements of the EBL should be considered upper limits. Some exceptions to this are the measurements from \citet{2024postmann}, or the dark-cloud technique from \citet{Mattila2012_darkcloud}. A complementary approach is based on measuring the flux from resolved sources, and extrapolating these data into galaxy counts for the whole galaxy population \citep[see, e.g.;][]{driver_measurements_2016,windhorst_jwst_2023}. This method gives an estimation of the lower limits of the EBL, since it does not account for diffuse, truly unresolved components.

As gamma rays propagate through the intergalactic medium, they can interact with EBL photons, leading to $e^-e^+$ pair production~\citep{1962ee_pair_creation,1967ee_pair_creation,1967ee_pair_creation2}, which has been observed in numerous systems \cite[e.g.,][]{2012Sci...338.1190A,Greaux_2024}. 
This process results in an energy-dependent attenuation of gamma-ray flux from extragalactic sources.
Higher-energy photons are more efficiently absorbed and the overall attenuation increases with redshift due to the longer propagation path through the photon field.
For instance, for sources at redshift $z=1$, the gamma-ray flux above $\sim30\,$\,GeV is attenuated. 
At lower values of $z$ this effect occurs at higher energies, of the order of hundreds of GeV or TeV.

An accurate characterization of the EBL is therefore necessary to infer the intrinsic spectra of distant sources from observations.
However, the uncertainties in EBL measurements and the variety of modeling methods have resulted in a range of EBL models with differing spectral shapes. Recent models generally converge within a narrow factor in the optical and infrared regimes, due to the use of similar datasets in their derivation.
A compilation of the most recent EBL models can be found in \citet{manuel_meyer_2022_ebltable}.

Most of the observed extragalactic VHE sources are blazars, a subclass of active galactic nuclei (AGN), that harbor relativistic jets closely oriented to the line of sight to Earth. 
The general approach by the gamma-ray community in the analysis of blazars is to use fixed EBL models (sometimes allowing for an overall renormalization) and compare the results for different models \citep[e.g.,][]{2013A&A...550A...4H,hessEBL,2019A&A...627A.110B,2019MNRAS.486.4233A,Abeysekara_2019}.
Another method is to simplify the EBL spectral shape to a sum of phenomenological functions \citep[e.g.,][]{2012A&A...542A..59M,2015biteau,Greaux_2024}.
In light of the increasing amount of data from current gamma-ray telescopes and the forthcoming Cherenkov Telescope Array Observatory (CTAO; \citet{2019scta.book.....C}), it is high time to provide the community with an open-source code that allows one to constrain the cosmological and physical parameters underlying the EBL. 

There are several methodologies to model the EBL.
Forward-evolution models calculate the EBL by following the formation and evolution of structure in the Universe. These approaches are built on hierarchical dark-matter halo formation, merger trees, and semi-analytical or hydrodynamical prescriptions for galaxy formation. The EBL is then obtained by integrating the modeled galaxy populations and their evolving spectral energy distributions across cosmic time \citep[see, e.g.,][]{2012MNRAS.423.1992S,Gilmore2012_SAM_EBL,2013ApJ...768..197I}.
Empirical models reconstruct the EBL based primarily on observations, so they depend very little on theoretical assumptions. These approaches use galaxy surveys to build luminosity functions, derive emissivities, and integrate them over redshift to obtain the EBL \citep[see, e.g.,][]{2017A&A...603A..34F,saldana-lopez_observational_2021}.
In this work, we focus on a phenomenological approach, initially developed by \citet{kneiske_implications_2002}, and subsequently used in other works \citep{2004A&A...413..807K,2010ApJ...712..238F, raue_probing_2012, finke_modeling_2022}. 
This approach calculates the EBL intensity by modeling the sources of radiation, using their emissivities and evolution through time.
Previously, we introduced a framework for modeling the COB \citep{2024Porras}, which we now extend to the CIB and thus the full EBL. 
Our work is accompanied by the release of the modeling software \textsc{niebla}, along with example use cases.\footnote{\url{https://github.com/saraporrasbedmar/niebla}\\
\textsc{niebla} means ``fog'' in Spanish which, similarly to the EBL, blocks distant observations.}
The code incorporates various sources contributing to the COB, such as stellar populations \citep[see, e.g.,][]{driver_measurements_2016,windhorst_jwst_2023}, intra-halo light (IHL) \citep[see, e.g.,][]{Purcell_2007} and stripped stars \citep[see, e.g.,][]{gotberg_y_impact_2019}.
The radiation from these sources is then absorbed by dust and reemitted in the infrared regime. 
We also include a possible additional contribution to the EBL from the decay of axion-like dark matter into photons.
Our code includes several common parametrizations for these processes but also accepts custom inputs provided by the user.

This work is structured as follows: in Section~\ref{sect:TheoreticalBackground} we introduce the theoretical background to the phenomenological modeling of the EBL, detailing the sources that contribute to its intensity. 
In Section~\ref{section:MeasurementsEBL} we introduce the observational data that we use, and in Section~\ref{sect:fitToData} we describe the fitting procedure we perform together with the results. 
A description of the functionality of \textsc{niebla} is provided in Section~\ref{sect:ModelRundown}. 
In Section~\ref{sect:UseCaseDustProperties} we present a use case for our EBL models. 
We simulate a VHE observation with the Large High Altitude Air Shower Observatory (LHAASO) of Mkn~501 assuming spectral parameters as inferred from the 1997 flare observed with the HEGRA Cherenkov telescopes \citep{1999A&A_FlareInitial,2001A&AFlareReanalysis}. 
Using this simulation, we demonstrate how it is possible to constrain cosmic dust properties using the different prescriptions of dust reemission included in our code. 
Finally, in Section~\ref{sect:conclusions}, we provide our conclusions.

\section{Theoretical background} \label{sect:TheoreticalBackground}

The EBL traces the integrated light produced over the course of cosmic history.
Its intensity can be calculated through
\begin{equation} \label{eq:nuInu_general}
    \nu I_{\nu}(\lambda,z) = \frac{c^2}{4\pi\lambda}\int\limits_{z}^{z_\mathrm{max}} \varepsilon_{\nu'}\left(\lambda\frac{1+z}{1+z'}, z'\right)\bigg | \frac{\mathrm{d}t}{\mathrm{d}z'} \bigg | \mathrm{d}z',
\end{equation}
with
\begin{equation}
\hspace{0.3cm}\frac{\mathrm{d}t}{\mathrm{d}z} =\frac{-1}{(1+z) H(z)},
\end{equation}
where we integrate the emissivity $\varepsilon_{\nu}(\lambda, z)$ of the source population over the time it has been emitting. The emissivity depends on wavelength, which is redshifted over cosmic time, and the redshift itself. The maximum redshift $z_\mathrm{max}$ defines the redshift at which the population started emitting.
The Hubble parameter is defined by the $\Lambda$CDM parameters as $H(z) = H_0 \sqrt{\Omega_{\Lambda} + \Omega_\mathrm{m} (1 + z)^3}$. We use the values $\Omega_\mathrm{m} = 0.3$, $\Omega_\mathrm{\Lambda} = 0.7$ and $H_0=70\,\mathrm{km\,s^{-1}\,Mpc^{-1}}$.

In general, the total emissivity is a sum of contributions of the COB and CIB, 
\begin{equation} \label{eq:sumCOBandCIBemiss}
    \varepsilon_{\nu}(\lambda, z) = f_\mathrm{esc, dust} \varepsilon_{\nu_\mathrm{COB}}(\lambda, z) + \varepsilon_{\nu_\mathrm{CIB}}(\lambda, z),
\end{equation}
where $f_\mathrm{esc, dust}$ is the fraction of light that avoids dust absorption. Possible parametrizations for $f_\mathrm{esc, dust}$ are given by, e.g., \citet{kneiske_implications_2002,razzaque_stellar_2009, finke_modeling_2022, the_fermi-lat_collaboration_gamma-ray_2018}. These are included in our code and are listed in Appendix~\ref{appendix:parametrizationsInCode}.
The emissivity of the CIB is then the star light absorbed in the interstellar medium by dust particles, and reemitted at infrared wavelengths.
We now provide details on the modeling of the COB and CIB.

\subsection{Optical regime}
The optical part of the EBL is dominated by star light.
Additional sources are the optical emission of AGN, the IHL, stripped stars, or more exotic contributions such as the decay of axion-like particles (ALPs). The total emissivity will be a sum of all these different contributions.

\subsubsection{Stellar contribution}

A simple stellar population (SSP) is a group of stars born from a single gas cloud.
The time scale for this creation is of the order of $10^7$ years, which can be approximated as instantaneous in cosmic terms. Following this first burst no more star formation occurs.
The emissivity of a stellar population can be calculated as
\begin{equation} \label{eq:emissStellar}
    \varepsilon_{\nu_\mathrm{stellar}}(\lambda, z) = \int^{z_\mathrm{max}}_{z}L_{\nu}^\mathrm{SSP}\left(\lambda, \tau_\star, Z\right)\,\rho_{\star}(z')\bigg|\frac{\mathrm{d}t'}{\mathrm{d}z'}\bigg|\mathrm{d}z',
\end{equation}
where $L^\mathrm{SSP}_{\nu}$ is the luminosity of the SSP at a given age and metallicity, the SSP age $\tau_\star$ can be written as $\tau_\star = t(z) - t(z')$, and $\rho_{\star}$ is the cosmic star formation rate density (SFRD).
The SFRD is the amount of stars born at a certain redshift per unit of volume. 
We assume that the metallicity of the SSP at a given redshift is equal to the mean metallicity of the Universe at that redshift $Z(z)$.
Popular parametrizations of $\rho_{\star}$ are provided by \citet{madau_cosmic_2014,haardt_radiative_2012,finke_modeling_2022} and for the mean metallicity evolution of the Universe by \citet{2022Tanikawa}. All are included in our code and are listed in Appendix~\ref{appendix:parametrizationsInCode}.

The SSP luminosities also depend on the adopted initial mass function (IMF). The IMF defines the relative number of stars formed at each stellar mass, and, together with the minimum and maximum stellar masses allowed in the SSP, it governs both its spectral shape and evolution with time.

Spectra of SSPs are extensively studied in the literature, and several computational tools for their numerical simulation have been developed. 
Examples of these codes are \textsc{Starburst99} \citep{leitherer_starburst99_1999,leitherer_effects_2014}, \textsc{Pégase} \citep{fioc_pegase3_2019}, and \textsc{PopStar} \citep{2009MNRAS.398..451M}.
They provide $L_{\lambda}(t_\star, Z)$, which serve as input for Eq.~\eqref{eq:emissStellar}.

An additional contribution could be stripped stars, which are usually not included in the SSP spectra. These stars have stripped their envelopes through the interaction with a binary companion. They emit a fraction of their radiation as ionizing photons, with energies larger than the ones predicted by stellar evolutionary tracks \citep{gotberg_y_impact_2019}.

\subsubsection{Intra-halo light}

The IHL is the light caused by stars that have been ejected from their galaxies by dynamical events \citep[see, e.g.,][]{Purcell_2007,galaxies9030060}. These stars would create a faint halo surrounding their host galaxies, but would not have been used to calculate the SFRD \citep[see, e.g.,][]{2013ApJ...770...57B,madau_cosmic_2014}. We follow the parametrization described by \citet{bernal_seeking_2022} to calculate the emissivity
\begin{equation} \label{eq:ihl_emiss}
     \varepsilon_{\lambda, \mathrm{IHL}} (\lambda, z) = \int ^{M_\mathrm{max}}_{M_\mathrm{min}} \frac{dn_\mathrm{h}}{dM} \, L_{\lambda, \mathrm{IHL}}(M, z) \mathrm{d}M,
\end{equation}
where $dn_\mathrm{h}/dM$ is the halo mass function and $L_{\lambda, \mathrm{IHL}}(M, z)$ is the IHL light specific luminosity emitted by a halo of mass $M$.
We use the same parameters as \citet{bernal_seeking_2022} to characterize $L_{\lambda, \mathrm{IHL}}(M, z)$, taking their definition and reported mean values for the fraction of the total halo luminosity coming from IHL and redshift power-law dependence.

\subsubsection{Cosmic ALP decay}

Axions are hypothetical pseudo-Nambu-Goldstone bosons, which are a consequence of the Peccei-Quinn solution to the strong CP problem in QCD~\citep{peccei_constraints_1977,peccei_cp_1977}.  
In contrast to axions, ALPs do not solve the strong CP problem and their mass $m_a$ is independent of the coupling constant to photons $g_{a\gamma}$.
Both axions and ALPs are candidates to make up cold dark matter (DM)~\citep[see, e.g.,][]{2012Arias}. 
If this were the case, there would be a continuous decay of axions or ALPs into two photons through cosmic time potentially contributing to the EBL intensity. 
The emissivity of cosmic ALP decay is given by~\citep{overduin_dark_2004}
\begin{equation} \label{eq:axionEmiss}
    \varepsilon_{\lambda, \mathrm{ALP}} (\lambda) \propto \exp\left(\frac{(\lambda - \lambda_\mathrm{decay})^2}{2\sigma_\mathrm{decay}^2}\right),
\end{equation}
where the wavelength of the decay photon is $\lambda_\mathrm{decay} = 2 h c~/ \left(m_a c^2\right)$, with the Planck constant $h$ and the speed of light in vacuum $c$.
The width of this Gaussian profile is $\sigma_\mathrm{decay} \lesssim 10^{-3}\lambda_\mathrm{decay}$ \citep{Vdisp1999, Vdisp2017, 2024Porras}. To compute its corresponding contribution to the EBL intensity, this narrow spectral feature can be approximated by a delta function. Substituting the approximation into Eq.~\eqref{eq:nuInu_general}, the resulting EBL intensity is
\begin{equation} \label{eq:axiondecayCOSMIC}
\nu I_{\nu} (\lambda,z) = \frac{\Omega_a\rho_\mathrm{crit, 0} c^4}{64 \pi} \frac{\left(m_a c^2\right)^2 g_{a\gamma}^2}{\lambda H(z_{\ast})} \Theta \left(\lambda -\frac{2 h c }{m_a c^2}\right),
\end{equation}
where $ z_{\ast} = \frac{m_a c^2}{2}\frac{\lambda}{h c}(1+z) - 1$, $\Omega_a$ is the fraction of DM in the form of axions and $\rho_\mathrm{crit, 0}$ is today's critical matter density of the Universe. 
A more detailed derivation can be found in \citet{2024Porras}.

\subsubsection{Other contributions}
The list of sources introduced above is not exhaustive. 
Additional contributions to the EBL could come, e.g., from AGN as their spectra span the entire electromagnetic spectrum~\citep[see, e.g.,][]{2017A&ARv..25....2P}. 
However their contribution to the EBL intensity is as low as $\sim10$\% of the stellar one \citep{10.1093/Andrews,10.1093Khaire}.
We do not include a parametrization for this source in \textsc{niebla}.
However, any extra contribution to the EBL can always be provided by the user either in the form of luminosities or emissivities. 
We provide tutorials in the \textsc{GitHub} repository on how to do this.\footnote{\url{https://github.com/saraporrasbedmar/niebla/tree/main/notebooks}}

\subsection{Dust reemission and the infrared}

The reemission of absorbed stellar light by dust is a key component of the CIB. \textsc{Niebla} provides two different approaches for modeling this infrared emission. The first uses spectral templates that describe the dust emission of galaxies or SSPs \citep{raue_probing_2012}. 
The second one uses combinations of black bodies to approximate emission from dust at different temperatures \citep[see, e.g.,][]{kneiske_implications_2002,finke_modeling_2022}. We proceed to describe both methods.

\subsubsection{Spectral templates}

In our first approach we use empirically calibrated spectral templates that describe the dust emission of galaxies or SSPs. There are two sets of templates already included in \textsc{niebla}, but custom ones can be provided by the user.

\begin{figure}[ht]
    \centering
    \includegraphics[width=0.95\linewidth]{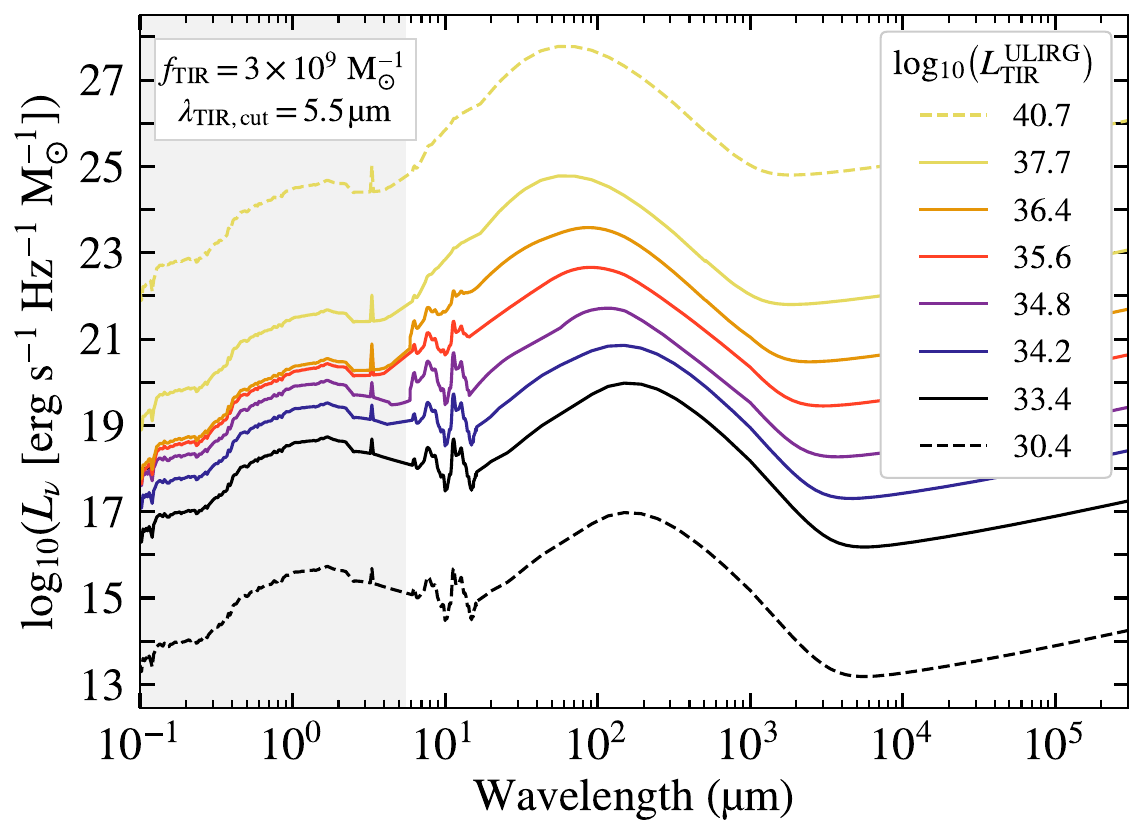}
    \caption{Set of synthetic spectra for dust reemission from ULIRGs \citep{chary2001}. We use $f_\mathrm{TIR} = 3\times10^9\,\mathrm{M}_{\odot}^{-1}$ to convert between galactic and SSP luminosities, and $\lambda_\mathrm{TIR, cut}=5.5$µm, represented by a shaded gray region below this value. We show the original spectra as solid lines, and our extrapolations as dashed lines.
} 
    \label{fig:chary2001spectra}
\end{figure}

Firstly, we include the infrared templates by \citet{chary2001}, hereafter \textit{Chary}, which are synthetic galactic infrared spectra derived from observations of ultraluminous infrared galaxies (ULIRGs).
These templates are normalized to the total infrared luminosity of the ULIRG systems $L_\mathrm{TIR}^\mathrm{ULIRG}$, so it is necessary to rescale them to match the expected infrared luminosity of our stellar model.
This rescaling is implemented through the parameter $f_\mathrm{TIR}$. 
Since the optical stellar contribution is already included in our EBL calculation, we follow  
the approach of \citet{raue_probing_2012}, and apply a wavelength cutoff in the dust luminosity templates below a certain $\lambda_\mathrm{TIR, cut}$.
The total infrared light $L_\mathrm{TIR}$ of a \textit{Chary} template is then calculated through 
\begin{equation} \label{eq:Ltir_int}
    L_\mathrm{TIR}^{Chary} = \frac{1}{f_\mathrm{TIR}} \int L_{\nu}^\mathrm{ULIRG}(\lambda) \Theta \left(\lambda -\lambda_\mathrm{TIR, cut}\right) \mathrm{d}\lambda.
\end{equation}
where $ L_{\nu}^\mathrm{ULIRG}$ are the original galactic templates.
In Fig.~\ref{fig:chary2001spectra}, we show examples of the rescaled \textit{Chary} templates using $f_\mathrm{TIR}=3\times10^9 \mathrm{M}_\odot^{-1}$ and $\lambda_\mathrm{TIR, cut}=5.5\,$µm, together with extrapolations beyond the minimum and maximum values of $L_\mathrm{TIR}$ provided by the templates. 
This extrapolation is performed by rescaling the limit cases without modifying the spectral shape.

\begin{figure}[htb]
    \centering
    \includegraphics[width=0.95\linewidth]{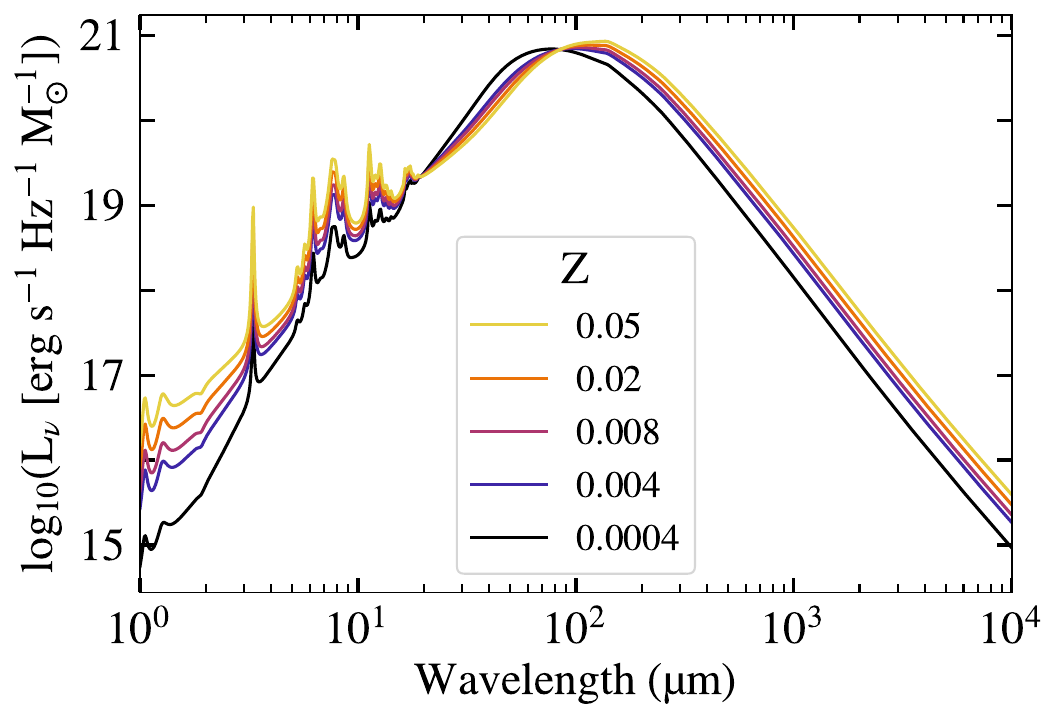}
    \caption{Synthetic spectra of dust emission in the infrared for different metallicities generated with the BOSA software \citep{2021BOSA}. The spectra are already normalized to $L_\mathrm{TIR} = 1\,\mathrm{L}_\odot$.}
    \label{fig:bosa_spectra}
\end{figure}

The second spectral library that we use are the BOSA templates \citep{2021BOSA}.
The BOSA project is not a static library but a computational tool that generates sets of spectral templates based on specified properties.
We use the BOSA templates calculated for different metallicities, see \citet{2021BOSA} for further details.
These templates are provided with their total infrared luminosity already normalized to the luminosity of a reference single stellar population, expressed in units of the solar luminosity $L_\odot$. Therefore, no additional normalization is required for our analysis.
In Fig.~\ref{fig:bosa_spectra}, we show the dust reemission spectra for different metallicities.\footnote{The BOSA package requires as an input the logarithm of the oxygen and hydrogen abundance, which we convert to metallicity using the formula $12 + \log(\mathrm{O/H}) = \log_{10}(Z/Z_\odot) + [\mathrm{O/H}]_\odot$ \citep{2020Zmeasurs, 2009Zsun}.}

To calculate the CIB emissivity from the dust templates, we need an additional ingredient, namely the integrated absorbed stellar luminosity, $L_\mathrm{abs, SSP}$, given by
\begin{equation} \label{eq:luminABS_int}
\begin{split}
    L_\mathrm{abs, SSP} (\tau_\star, Z) = \int &\left( 1 - f_\mathrm{esc, dust}(\lambda, z)\right) \\
    &\times L_{\nu}^\mathrm{SSP}\left(\lambda, \tau_\star, Z\right) \mathrm{d}\lambda.
    \end{split}
\end{equation}
We focus on stellar luminosities, since they dominate in the optical regime. Additionally, IHL is by definition emitted in regions outside of galaxies, so dust reemission templates do not account for it. We do not include contributions from ALPs, stripped stars or other sources.

We choose the spectral template that matches metallicity and $L_\mathrm{TIR}$ to the absorbed luminosity. The luminosity reemitted by dust is then  given by
\begin{equation}
    L_{\nu}^\mathrm{dust} = L_{\nu}^\mathrm{template}(Z, L_\mathrm{TIR}=L_\mathrm{abs, SSP} (\tau_\star, Z))
\end{equation}
where the $template$ can be \textit{Chary}, BOSA, or a custom one. We note that the $L_\mathrm{TIR}^{Chary}$ templates have two extra free parameters, as stated in Eq.~\eqref{eq:Ltir_int}.

The emissivity of dust reemission is complementary to that of the stellar emissivity in Eq.~\eqref{eq:emissStellar}, and can be expressed as
\begin{equation} \label{eq:emissStellarABS}
\begin{split}
    \varepsilon_{\nu_\mathrm{, CIB}}(\lambda, z) = \int^{z_\mathrm{max}}_{z} L_{\nu}^\mathrm{dust}\,\rho_{\star}(z')\bigg|\frac{\mathrm{d}t'}{\mathrm{d}z'}\bigg|\mathrm{d}z'.
\end{split}
\end{equation}
Note that with this definition, the factor $\left( 1 - f_\mathrm{esc, dust}(\lambda, z)\right)$ that seems to be missing in Eq.~\eqref{eq:sumCOBandCIBemiss} is introduced in the dust emissivity with Eq.~\eqref{eq:luminABS_int}.

The spectral templates may depend on the IMF assumed during their calibration. 
In \citet{chary2001}, a Salpeter IMF \citep{1955ApJ...121..161S} is adopted as part of the observational calibration. However, the spectral shapes of the templates is independent of this IMF calibration, and are therefore purely empirical.
In contrast, the BOSA library is explicitly constructed using the IMF for single objects from \citet{2003PASP..115..763C}. Consequently, when using the BOSA templates, the IMF of the SSP must be consistent with the Chabrier IMF.
We have added a Warning in the code when introducing dust reemission, so users can be aware of this issue. We also mention it in the notebooks we provide as documentation.

\subsubsection{Blackbody spectra}

The second method to calculate the infrared emission assumes different dust populations, where the reemitted light follows a blackbody spectrum for each population peaking at different temperatures.
For example, polycyclic aromatic hydrocarbon and very small grains are heated in galaxies to $T \sim \mathcal{O}(100\,\mathrm{K})$, and therefore emit in the mid infrared (MIR) \citep{chary2001,finke_modeling_2022}.
At higher wavelengths, cold dust is heated by interstellar radiation fields in galaxies and emits predominantly in the far infrared (FIR).

The emissivity of a black body, normalized over the frequency range, is given by
\begin{equation} \label{eq:blackbody1}
    \varepsilon_{\mathrm{BB,\,norm}}(\nu, T) = \frac{15}{\pi^4} \frac{1}{\nu} \frac{x^4}{e^{x} - 1} \hspace{0.5cm} \mathrm{with}\,\, x = \frac{h\nu}{k_\mathrm{B} T}.
\end{equation}
Here $k_\mathrm{B}$ is the Boltzmann constant and $T$ the temperature of the black body.
Previous EBL models, such as \citet{finke_modeling_2022}, have modeled the CIB as the sum of three black bodies, with one population emitting mainly in the MIR and two peaking at FIR wavelengths. The study from \citet{2019JCAP...04..043D} introduces one to three black bodies and compared different configurations.
We therefore also allow for multiple populations at associated temperatures $T_i$ and fraction of absorbed emissivity $f_i$. The total dust emission is then
\begin{align} \label{eq:dustReem_BB}
   \varepsilon_{\nu_\mathrm{CIB}}^{\mathrm{grey\,bodies}}(\lambda, z) &=  
 \sum_i  f_i \varepsilon_{\mathrm{BB,\,norm}}(\nu, T_i)\\
    &\times \int \left(1 - f_\mathrm{esc, dust}\right) \varepsilon_{\nu_\mathrm{COB}}(\lambda, z) \mathrm{d}\lambda, \nonumber
\end{align}
where we multiply the normalized blackbody spectra with the integrated emissivity that dust absorbs from the COB. The fractions $f_i$ are such that $\sum_if_i = 1$.

\section{Measurements of the EBL} \label{section:MeasurementsEBL}

Broadly speaking, there are two types of EBL measurements: lower and upper limits. Upper limits are derived from direct observations, where all foreground contributions are subtracted. Foregrounds are non isotropic contributions generated inside our Galaxy, and there is an active effort in the community to isolate the EBL from them.
Some of these foregrounds are the galactic dust and gas, atmosphere emission, and Zodiacal light \citep{matsumoto_reanalysis_2015,matsuura_new_2017}.
The New Horizons probe has claimed a tentative measurement \citep{2024postmann}, but the analysis still depends on subtraction of galactic foregrounds, so we also treat it as an upper limit here.
We show the upper limits as empty gray symbols in Fig.~\ref{fig:ebl_fits}. 
We do not use these data in our analysis, since it is uncertain whether all  foregrounds are fully subtracted.
Lower limits are based on galaxy counts \citep{driver_measurements_2016,windhorst_jwst_2023};
from the measured intensity, number, and mass of resolved galaxies one  extrapolates to the total contribution of the entire galactic population to the EBL intensity.
The lower limits are shown as filled data points in Fig.~\ref{fig:ebl_fits}.

In addition to the galaxy counts, we use emissivity data from galaxy surveys compiled by \citet{finke_modeling_2022} and a selection is shown in  Fig.~\ref{fig:emissivities} to further constrain our COB model. 
Furthermore, we use SFRD data compiled in \citet{the_fermi-lat_collaboration_gamma-ray_2018} which is derived from, e.g., luminosity measurements of galaxies in the ultraviolet and infrared, or using the nebular line emission that characterizes star formation \citep{madau_cosmic_2014}.
We show the SFRD data in Fig.~\ref{fig:sfr}.
Lastly, we also use measurements of the mean metallicity of the Universe as a function of redshift. These measurements are taken from \citet{2020Zmeasurs}, and we use them to determine the metallicity evolution of our SSPs and dust reemission templates. We show them in Fig.~\ref{fig:metallEvol}.

\section{Data analysis and results} \label{sect:fitToData}

We are now in the position to fit our EBL models for the different dust prescriptions to the observational data.
The best-fit parameters are determined through a combined $\chi^2$ for the different data sets,
\begin{equation} \label{eq:chi2}
\begin{split}
    \chi^2 =& \sum_i \left(\frac{\nu I_{\nu_\mathrm{model}} (\lambda_i) - \nu I_{\nu, i} }{\sigma_i}\right)^2 \\
    &+ \sum_i \left(\frac{\varepsilon_\mathrm{model} (\lambda_i, z) - \varepsilon_i}{\sigma_i}\right)^2 \\
    &+ \sum_i \left(\frac{\rho_{\star, \mathrm{model}} (z_i) - \rho_{\star, i} }{\sigma_i}\right)^2 \\
    &+ \sum_i \left(\frac{Z_{\mathrm{model}} (z_i) - Z_{i} }{\sigma_i}\right)^2,
\end{split}
\end{equation}
where $I_{\nu, i}$, $\varepsilon_i$, $\rho_{\star, i}$, and $Z_{i}$ denote the data for the galaxy number counts, emissivities, SFRD and metallicities, respectively, with corresponding uncertainties denoted with $\sigma_i$.
We perform two sets of fits. In the first one, we take the uncertainties of the data points reported in the literature at face value and use them as the $\sigma_i$ values in Eq.~\eqref{eq:chi2}.
In the second one, we add a fixed systematic error to the reported uncertainty in quadrature, expressed as a percentage of the measured value.
This additional term accounts for systematic offsets between the data sets from different surveys and methodologies, which in some cases are not in agreement within the $1\,\sigma$ confidence level. We test several values for this fractional uncertainty, and present the results for the value that yields a reduced $\chi^2$ closest to unity for all EBL models.
This is particularly evident in Figs.~\ref{fig:ebl_fits} and \ref{fig:sfr}.
In total, we have 534 data points in our fit, which is implemented with the least squares method from \textsc{iminuit} \citep{iminuit}.

From the list of sources of the COB we introduced in Section~\ref{sect:TheoreticalBackground}, we only use SSPs for our baseline models.
Other contributions are sub-dominant; stripped stars contribute at most 1\% of the SSP intensity at around $\lambda\sim0.3\,$µm, and the IHL intensity  is $\lesssim 10\%$ of the EBL intensity of our assumed stellar population peaking at $\lambda\sim5\,$µm, see \citet{2024Porras} for further details.

To generate our SSPs, we use the \textsc{Starburst99} code and assume a Kroupa IMF, with $M_\mathrm{min}=0.1\,\mathrm{M}_\odot$ and $M_\mathrm{max}=120\,\mathrm{M}_\odot$~\citep{kroupa2002}, and using the evolutionary Padova tracks.
In the BOSA model, we encounter a discrepancy in the IMF, as previously discussed in Section~\ref{sect:TheoreticalBackground}. These dust reemission templates assume a Chabrier IMF, which is formally different from the Kroupa IMF used for the SSPs. However, the differences between these IMFs are minimal across the relevant stellar mass range for our SSP templates, their shapes mostly overlap.
We take the prescription from \citet{madau_cosmic_2014} to parametrize the SFRD.
The SFRD measurements from \citet{the_fermi-lat_collaboration_gamma-ray_2018} are based on a Chabrier IMF. To convert the SFRD from Chabrier to Kroupa IMF, a constant factor of 0.94 must be applied \citep{madau_cosmic_2014}. However, in our analysis, we do not apply this rescaling explicitly, as we treat the normalization of our SFRD model as a free parameter during the fitting process.
The metallicity evolution is taken from \citet{2022Tanikawa}.
We assume the same functional form for the dust absorption as \citet{finke_modeling_2022}.
The mathematical formulas for these models are compiled in Appendix~\ref{appendix:parametrizationsInCode}.

Each of the dust reemission models has a different number of free parameters. We introduce two free parameters for \textit{Chary}, $f_\mathrm{TIR}$ and $\lambda_\mathrm{TIR, cut}$.
The BOSA templates have no extra parameters.
The number of the free parameters in the blackbody model depends on the number of assumed dust populations. 
We have decided to test a model with two black bodies (2BB) and fix one of those temperatures to $T_1=450\,$K to emulate the MIR dust emission.
Therefore, the free parameters in this case are one of the fractions $f_i$ and the second temperature $T_2$. We base these assumptions on the EBL modeling from \citet{finke_modeling_2022}.

\subsection*{Results using different dust reemission models}  \label{subsection:ResultsDiffDustModels}

\begin{figure*} [htb]
    \centering
    \includegraphics[width=0.95\linewidth]{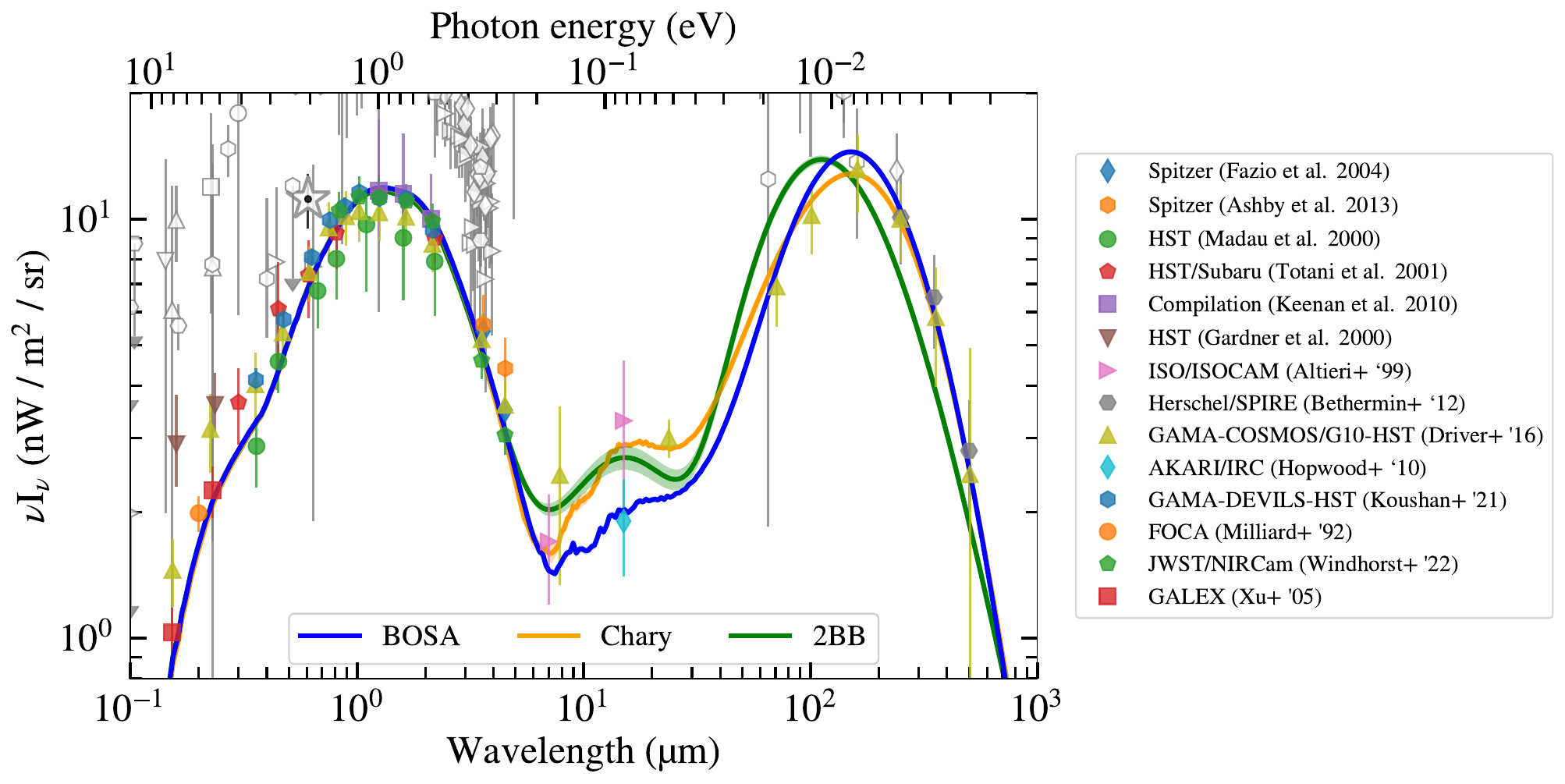}
    \caption{Overview of the EBL measurements used in this analysis. Data have been taken from \citet{hill_spectrum_2018} and \citet{Biteau_The_MM_EGAL_spectrum}. Upper limits of the EBL are represented as gray open symbols, lower limits as colored filled symbols. More detailed description of the measurements is given in Section~\ref{section:MeasurementsEBL}. 
    The best-fit results of our EBL models with three different prescriptions for the dust reemission are also displayed, see Section~\ref{subsection:ResultsDiffDustModels}.}
    \label{fig:ebl_fits}
\end{figure*}

\begin{table*}[htb]
    \centering
    \begin{tabular}{cccc} \midrule\midrule
        Parameter & BOSA & \textit{Chary} & 2BB \\ \midrule
        \multicolumn{4}{c}{$\rho_{\star}$} \\ \midrule
        $x_0$ $\left( \times 10^{-2} \,\mathrm{M}_{\odot}\,\mathrm{yr}^{-1}\,\mathrm{Mpc}^{-3}\right)$ & $1.38 \pm 0.06$ & $1.26\pm 0.05$ & $1.22\pm 0.05$ \\
        $x_1$ & $2.54\pm 0.09$  & $2.67\pm 0.09$ & $2.66\pm 0.09$\\
        $x_2$ & $3.03\pm 0.08$  & $3.00\pm 0.08$ & $3.03\pm 0.08$\\
        $x_3$ &  $6.3\pm0.1$  & $6.4\pm 0.1$ & $6.4\pm 0.1$\\ \midrule
        \multicolumn{4}{c}{\textit{Z}} \\ \midrule
        $a_0$ & $0.22\pm0.06$  & $0.20\pm 0.04$ & $-0.08\pm 0.06$\\
        $a_1$ &  $0.42\pm0.05$  & $0.44\pm 0.05$ & $0.23\pm 0.06$\\
        $a_2$ & $0.67\pm0.07$  & $0.61\pm 0.07$ 66& $0.9\pm 0.2$\\ \midrule
        \multicolumn{4}{c}{Dust reemission} \\ \midrule
        $f_\mathrm{TIR}$ $\left(\mathrm{M}_\odot^{-1}\right)$ (dex) & - &  $9.39\pm 0.04$ & - \\
        $\lambda_\mathrm{TIR, cut}$ (µm) & - &  $5.29\pm 0.12$ & - \\
        $f_1$ & - & - & $0.16\pm 0.01$ \\
        $T_2$ (K) & - & - & $64\pm 2$ \\
        \midrule
    \end{tabular}
    \caption{Best-fit parameters for our EBL models, detailed in Section~\ref{sect:fitToData} for the fit including a 14\,\% systematic uncertainty.
    See Section~\ref{sect:fitToData} for details.}
    \label{tab:bestfits}
\end{table*}

\begin{figure}[htb]
    \centering
    \includegraphics[width=0.99\linewidth]{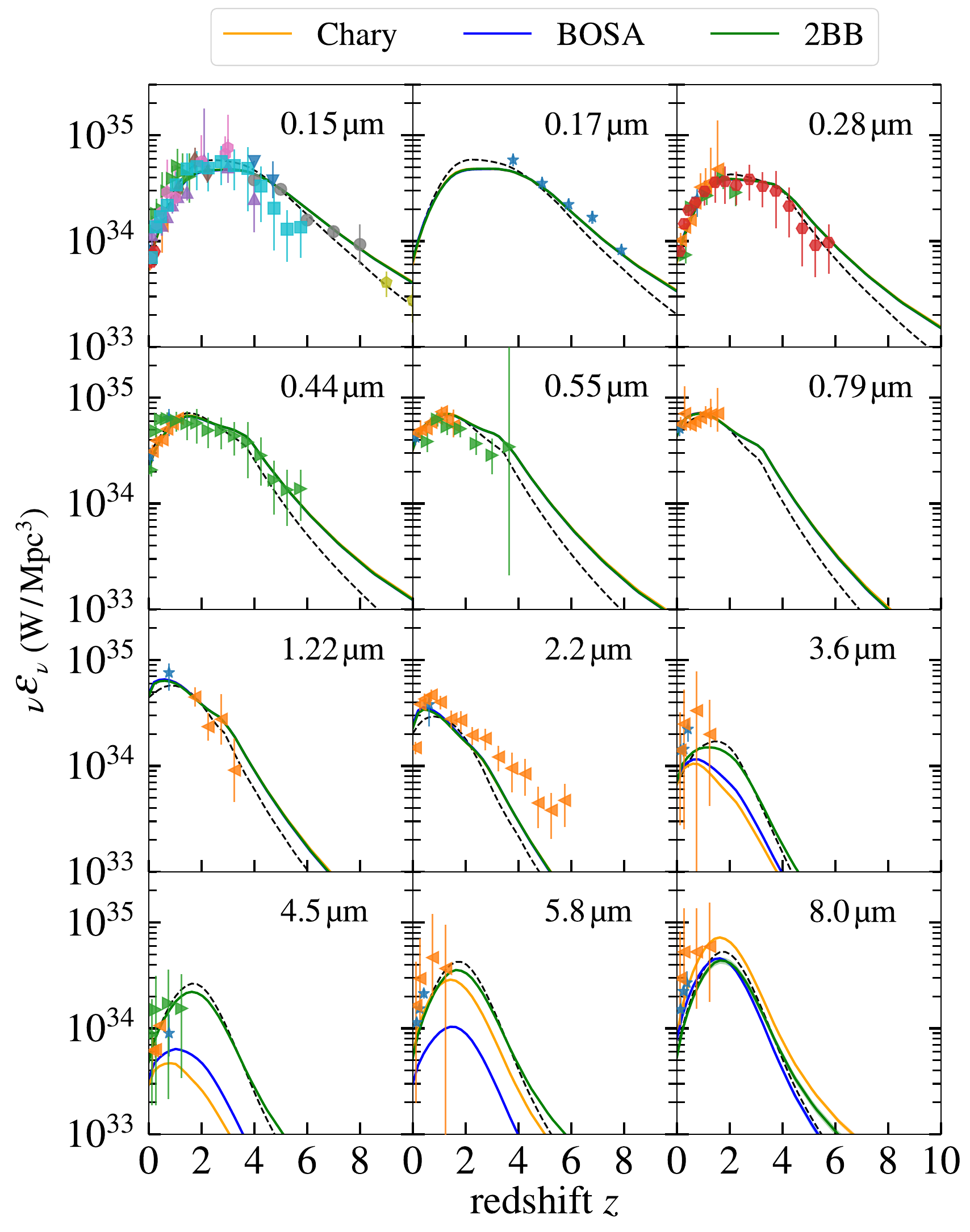}
    \caption{Emissivity data for specific wavelengths as a function of redshift. Observational data is taken from the compilation by \citet{finke_modeling_2022}. Solid lines are our best-fit curves for each of our models. The $1\,\sigma$ uncertainty band is too thin to be seen. Model A from \citet{finke_modeling_2022} is shown as dashed black lines for comparison.}
    \label{fig:emissivities}
\end{figure}

\begin{figure}[htb]
    \centering
    \includegraphics[width=0.99\linewidth]{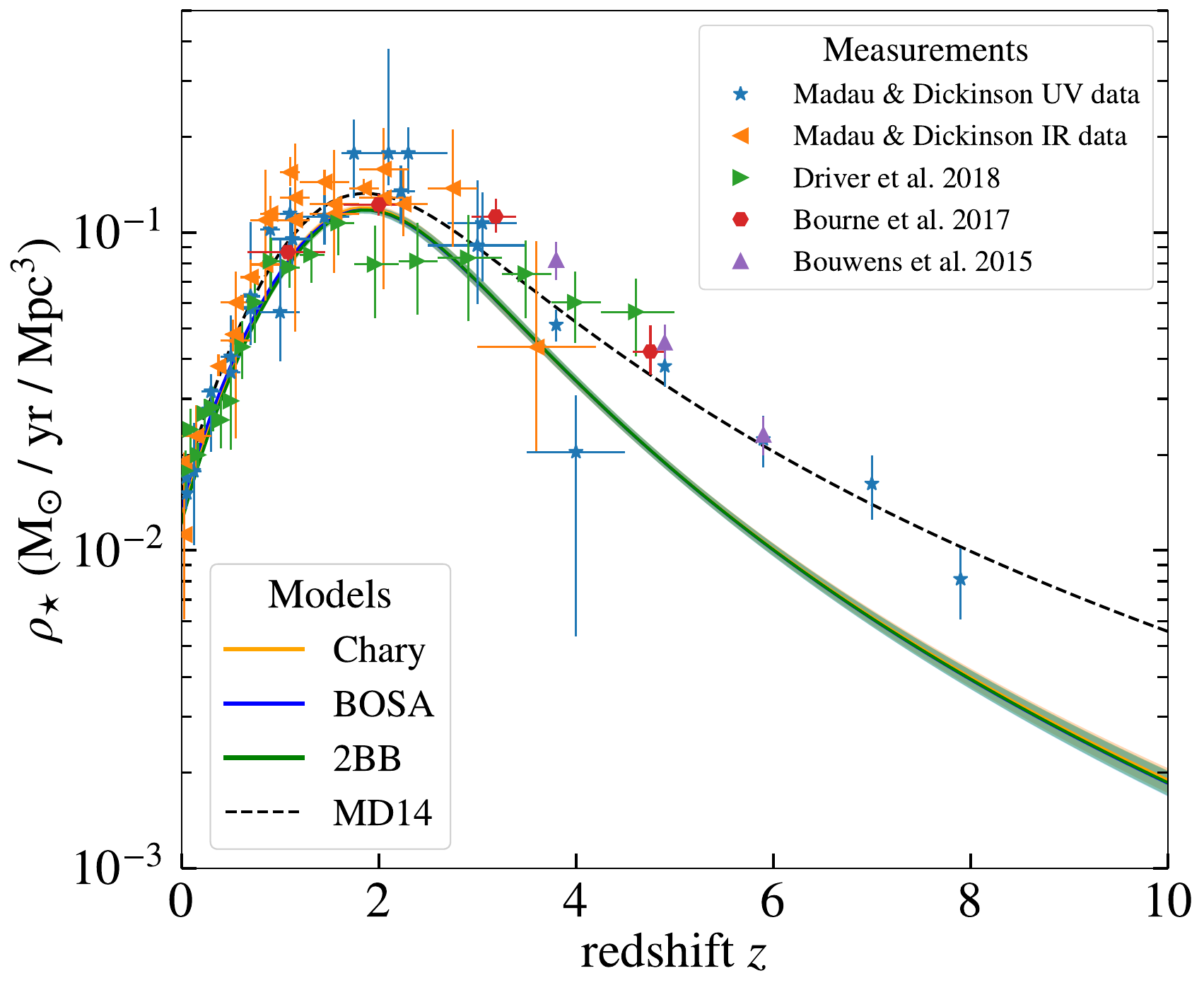}
    \caption{The SFRD as a function of redshift, with the $1\,\sigma$ uncertainty shown as light bands for each dust reemission model. Note that the bands overlap. Observational data compiled by \citet{the_fermi-lat_collaboration_gamma-ray_2018} are shown with different symbols. The parametrization of \citet{madau_cosmic_2014} (MD14 in the legend) is shown as a dashed line for comparison. 
    }
    \label{fig:sfr}
\end{figure}

\begin{figure}[htb]
    \centering
    \includegraphics[width=0.99\linewidth]{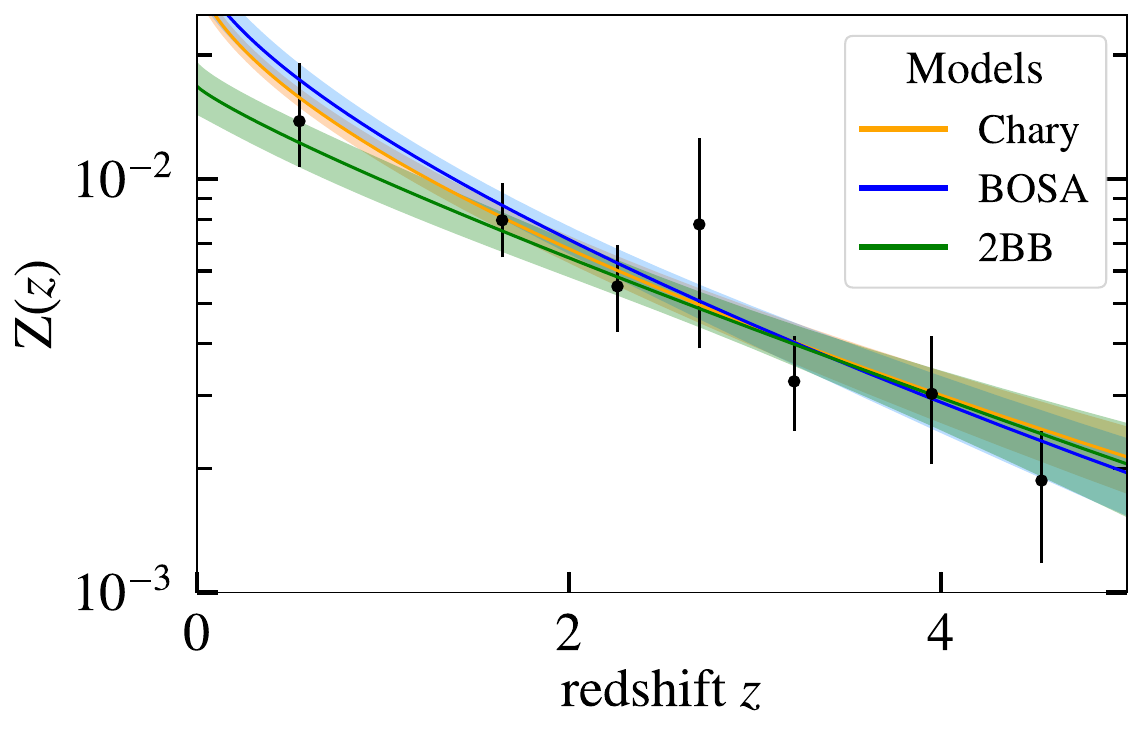}
    \caption{The mean metallicity \textit{Z} as a function of redshift. Observational data is taken from \citet{2020Zmeasurs}. The best-fit line and $1\,\sigma$ uncertainty are shown as a line and shaded region for each EBL model with a different dust reemission.
    }
    \label{fig:metallEvol}
\end{figure}

Running the fit procedure with several values for the fractional systematic uncertainty, we find that 14~\% is the value that yields a reduced $\chi^2$ closest to unity. The resulting best-fit parameters of this case are presented in Table~\ref{tab:bestfits}, while the $\chi^2$ values, split by individual data sets, are provided in Table~\ref{tab:chi2_various}.
This table also includes the resulting $\chi^2$ values computed without incorporating systematic uncertainties. The corresponding best-fit parameters for this case are provided in Appendix~\ref{appendix:correlation_matrices}. Generally, the best-fit parameters obtained with and without the inclusion of systematic uncertainties differ only in the second significant digit.
The following results, unless stated otherwise, refer to the fits performed with the additional 14\,\% systematic uncertainty included.
We show the results from the fits with the three dust reemission models in Fig.~\ref{fig:ebl_fits} for the EBL intensity, in Fig.~\ref{fig:emissivities} for our selection of the emissivities, in Fig.~\ref{fig:sfr} for SFRD, and Fig.~\ref{fig:metallEvol} for the metallicity.
The main contribution to the $\chi^2$ values comes from the emissivities in the optical regime, while the metallicity data have only marginal impact.
Among our models, the one with the dust reemission given by the \textit{Chary} templates has the smallest reduced $\chi^2$ value.
The full correlation matrices for all three models are provided in Appendix~\ref{appendix:correlation_matrices} for the fits including and not the 14~\% systematic uncertainty.
Correlation values remain similar in the cases with and without systematic uncertainty. In all scenarios, two clusters of correlated and anti-correlated parameters emerge within the SFRD and metallicity parameters. The correlation across these two groups are generally moderate. In contrast, the parameters related to dust reemission, $f_\mathrm{TIR}$, $\lambda_\mathrm{TIR, cut}$, $f_1$ and $T_2$, display only weak correlations with the rest of the parameter set, therefore being well constrained independently.

In Fig.~\ref{fig:emissivities} we compare our emissivities to those of Model A from~\citet{finke_modeling_2022}, shown as dashed black lines.
This model yields results similar to our 2BB model in the MIR (corresponding to the highest wavelengths $\geqslant 4.5$\,µm shown in the bottom row of Fig.~\ref{fig:emissivities}), which is expected given that both models utilize the same methodology for dust reemission and the same temperature for the black body in this regime.

In Appendix~\ref{appendix:EBL_figure} we provide the redshift evolution of our EBL models, and compare these with established benchmark models.

All three methodologies of dust reemission produce acceptable fits within observational uncertainties. 
Nevertheless, they differ in their physical interpretations.
In the \textit{Chary} and BOSA models, the infrared spectral shape is fixed to that of the dust reemission templates, so the fit adjusts the global emissivity evolution. In contrast, the 2BB model includes explicit temperatures and fractional contributions of distinct dust components, allowing the infrared EBL shape to vary more freely. As a result, the fitted parameters mainly reflect how each dust methodology redistributes absorbed light into the infrared, while the optical component remains robust across all models.
Importantly, the best-fit parameter values in all cases fall within the ranges observed in typical galaxies; the implied dust temperatures $T_i$ are consistent with measurements of local galaxies \citep[see, e.g.,][]{1990A&A...237..215D,Dwek_1997}, and $f_\mathrm{TIR}$ should be of the order of the stellar mass of a galaxy \citep{raue_probing_2012}.

We also tried to fit the blackbody reemission with three blackbody components instead of two, fixing two temperatures $T_1=450\,$K and $T_2=70\,$K and leaving two of the $f_i$ as free parameters. 
However, the fit preferred  $f_2 = 0$, removing the need for this additional contribution.

\begin{table*}
    \centering
    \begin{tabular}{ccccccccccc}\midrule\midrule
    & \multicolumn{5}{c}{14\% systematics} & \multicolumn{5}{c}{No systematics added} \\
    \cmidrule(lr){2-6}\cmidrule(lr){7-11}
        & $\chi^2_\mathrm{EBL}$ & $\chi^2_\varepsilon$ & $\chi^2_{\rho_{\star}}$ & $\chi^2_Z$ & $\chi^2_\mathrm{Total}/\mathrm{d.o.f.}$ &  $\chi^2_\mathrm{EBL}$ & $\chi^2_\varepsilon$ & $\chi^2_{\rho_{\star}}$ & $\chi^2_Z$ & $\chi^2_\mathrm{Total}/\mathrm{d.o.f.}$ \\ \midrule
       $\#$ data points & 68 & 384 & 75 & 7 & & 68 & 384 & 75 & 7  \\
       \midrule
        BOSA & 37.64 & 376 & 113.2 & 2.34 & $528.8 / 527 = 1.00$ & 75.47 & 1216 & 226.0 & 3.93 & $1521 / 527 = 2.9$ \\
        \textit{Chary} & 39.30 & 350 & 115.8 &  2.08 & $ 506.7 / 525 = 0.97 $  & 73.66 & 1029 & 256.6 & 9.03 & $ 1368 / 525 = 2.6 $\\
        2BB & 48.86 & 389 & 122.5 & 1.72 & $561.7 / 525 = 1.07 $  & 83.98 & 988 & 278.1 & 11.97 & $1362 / 525 = 2.6 $\\\midrule
    \end{tabular}
    \caption{$\chi^2$ values for the dust reemission models described in the text, given for each of the expressions in the fit, as well as the total and reduced $\chi^2$ value. We also list the number of data points for each of the contributions to the $\chi^2$. We provide the values for the fits with and without an additional 14\,\% of  systematic uncertainty. See Section~\ref{sect:fitToData} for details.}
    \label{tab:chi2_various}
\end{table*}

\section{Model rundown} \label{sect:ModelRundown}

\textsc{Niebla} is a package that aims to balance between memory consumption and computational efficiency, prioritizing efficiency for fitting procedures. It stores the intermediate values necessary to efficiently recalculate the outputs when input parameters are modified.
The model requires $t\sim3\,$s to compute the COB and $t\leq10\,$s to compute the entire EBL, depending on the dust reemission model. The primary factor for this time span is the evaluation of splines required in the calculations.
A fit with \textsc{iminuit} for 1-2 free parameters typically converges in $t\sim20\,$mins. The fits presented in Section~\ref{sect:fitToData} have $7-9$ free parameters, which converge in a span of $t\sim1-3\,$h. The fastest EBL to converge is 2BB, because it uses less splines than the spectral templates. The typical memory usage of the code is approximately $600$\,MB. To obtain these results we have used a personal computer.

We provide \textsc{Jupyter Notebook} scripts\footnote{\url{https://github.com/saraporrasbedmar/niebla/tree/main/notebooks}} that introduce all the necessary inputs for the code and explain how to obtain the various outputs. We provide a list of the parametrizations and inputs currently present in the code in Appendix~\ref{appendix:parametrizationsInCode}. All parametrizations and inputs accept custom functions or spectra, which can be supplied either by listing new directories, coding new functions, or by using splines. All the necessary information to customize inputs is included in the \textsc{notebooks}.

By default, the code adopts a standard $\Lambda$CDM cosmology with parameters $H_0 = 70\,\mathrm{km}\,\mathrm{s}^{-1}\,\mathrm{Mpc}^{-1}$, $\Omega_\mathrm{matter}=0.3$, $\Omega_\Lambda=0.7$ and $\Omega_\mathrm{baryons}=0.0453$. 
The values of these parameters can be varied, and other cosmological models can also be introduced. We detail this procedure and discuss caveats in Appendix~\ref{appendix:cosmology}.

We briefly outline the implementation of the synthetic templates in the code. The required quantities are obtained through spline interpolation. The synthetic luminosity spectra of SSPs depend on wavelength, age, and metallicity. For wavelengths and ages outside the range covered by the templates, the splines are set to return zero. For metallicities beyond the available range, the spectra are extrapolated by adopting the spectral shape corresponding to the nearest metallicity template.
For the dust reemission libraries we adopt a similar approach. 
We do not perform extrapolations for wavelength values outside the template ranges but extrapolate for values of $L_\mathrm{TIR}$ outside the templates, where the same spectral shape as the closest value with the requested $L_\mathrm{TIR}$ is returned. 
We provide an example of this approach in Fig.~\ref{fig:chary2001spectra}, where the dashed lines show our extrapolation.
The \textit{Chary} library does not incorporate metallicity dependence, so luminosity values remain unchanged with metallicity evolution. 
BOSA templates return the same spectral shape as the closest metallicity available when the requested metallicity values are outside of the provided range.

The code allows as many black bodies as the user may want to add. 
The user needs to provide (initial) values of all $T_i$ with $i=1, 2, \dots n$ and the corresponding $f_i$ for the first $n-1$ terms, where $f_n$ is calculated from the condition $\sum_i f_i = 1$.

The IHL calculation uses a \textsc{C++} library to calculate the $dn_\mathrm{h}/dM$ from Eq.~\eqref{eq:ihl_emiss} called \textsc{hmf} \citep{murray_hmfcalc_2013}.
For users not interested in the IHL contribution, it is possible (and recommended) to install \textsc{niebla} without this dependency.

The stripped stars follow the same calculation as the general SSP calculations, but with different luminosity spectra. 
Here, we used the spectra listed in the \textsc{Starburst99} webpage.\footnote{\url{http://www.stsci.edu/science/starburst99/data/published_runs_starburst99.zip}}

The cosmic ALP contribution is slightly more delicate to compute than the other EBL sources.
Due to the nature of ALP decay, there is a sharp cutoff at the wavelength related to the ALP mass.
Splines are therefore discouraged at any step of the calculation, as they generally fail to reproduce the spectrum if the input grid does not have the necessary resolution to resolve the cutoff.
Our recommendation is to calculate the ALP decay intensity for each wavelength and redshift needed.
This takes of the order of $t\sim0.01\,$s as we provide the analytical expression for the EBL intensity caused by ALP decay (see Eq.~\eqref{eq:axiondecayCOSMIC}).

\section{Use case of \textsc{niebla}: determining dust properties} \label{sect:UseCaseDustProperties}

In this Section, we present a use case for \textsc{niebla}. We simulate observations from a VHE source and compare the results obtained with the different EBL models we have calculated. Our objective is to assess whether it is possible to differentiate between the dust reemission models with future VHE observations. Since the models are not nested, unlike in \citet{2019JCAP...04..043D}, a simple likelihood ratio is not appropriate for this comparison. We now introduce the pipeline that we follow for this analysis.

As can be seen from Figs.~\ref{fig:ebl_fits} and \ref{fig:emissivities}, data in the mid- and far infrared regime is relatively scarce in comparison to optical data.
As discussed in Section~\ref{sec:Intro}, VHE gamma rays from blazars offer an alternative method to probe the EBL intensity through their absorption on the EBL. 
The observed photon flux $\phi_\mathrm{obs}$ is related to the intrinsic spectrum $\phi_\mathrm{int}$ emitted by the blazar through $\phi_\mathrm{obs} = \exp(-\tau) \phi_\mathrm{int}$, where $\tau$ is the optical depth given by
\begin{equation} \label{eq:tau}
\begin{split}
        \tau (E_0, z_0) = &\int_0 ^{z_0} \mathrm{d}z \frac{\mathrm{d} L}{\mathrm{d} z}(z) \int_0 ^{\infty} \mathrm{d}\epsilon \frac{\mathrm{d} n}{\mathrm{d} \epsilon} (\epsilon, z) \\
        & \int_{-1} ^{1} \mathrm{d}\mu \frac{1 - \mu}{2} \sigma_{\gamma \gamma} \left[\beta\left(E_0, z, \epsilon, \mu\right)\right],
\end{split}
\end{equation}
where $\epsilon=h\nu$ is the EBL photon energy, $\mathrm{d}n/\mathrm{d}\epsilon$ is the EBL photon density which is related to the EBL intensity through $\mathrm{d}n/\mathrm{d}\epsilon = c /(4 \pi) \epsilon^2 \, \nu I_{\nu}$. The distance element is $\mathrm{d} L/\mathrm{d} z = c / (1+z)/H(z)$, $\mu$ is the cosine of the angle between the gamma ray and the EBL photon, and $\sigma_{\gamma \gamma}$ is the pair production cross section, which depends on the parameter $\beta$ as \citep[see, e.g.,][]{2015biteau}
\begin{equation}
    \sigma_{\gamma \gamma} = \frac{3\sigma_\mathrm{T}}{16} \left(1-\beta^2\right) \Bigg[ -4\beta + 2\beta^3 + \left(3-\beta^4\right)\ln{\frac{1+\beta}{1-\beta}}\Bigg], 
\end{equation}
where $\sigma_\mathrm{T}$ is the Thomson cross-section, and $\beta$ is defined as
\begin{equation}
    \beta^2 = 1 - \frac{2 m_\mathrm{e} c^4}{E_0 \epsilon} \frac{1}{1+z}\frac{1}{1 - \mu},
\end{equation}
where $m_\mathrm{e}$ is the mass of the electron.

For a gamma ray at energy $E$, the pair production cross section peaks at an EBL wavelength of $\lambda \simeq 1.24 (E / 1\,\mathrm{TeV})\,$µm \citep{2008A&A...487..837F}. 
Therefore, if we want to constrain dust properties, observations beyond tens of TeV are necessary. 
For this reason, we investigate whether we would distinguish our dust reemission models using future observations of Mkn~501, a close-by blazar located at a redshift of $z= 0.034$.
This source exhibited a major flare in 1997, reaching energies of $\sim 20\,$TeV, which was observed with HEGRA, an array of imaging air Cherenkov telescopes \citep{1999A&A_FlareInitial}.
We assume here that the source undergoes a similar major outbreak that is observed with the LHAASO air shower array.

\begin{figure*}[htb]
    \centering
    \includegraphics[width=0.95\linewidth]{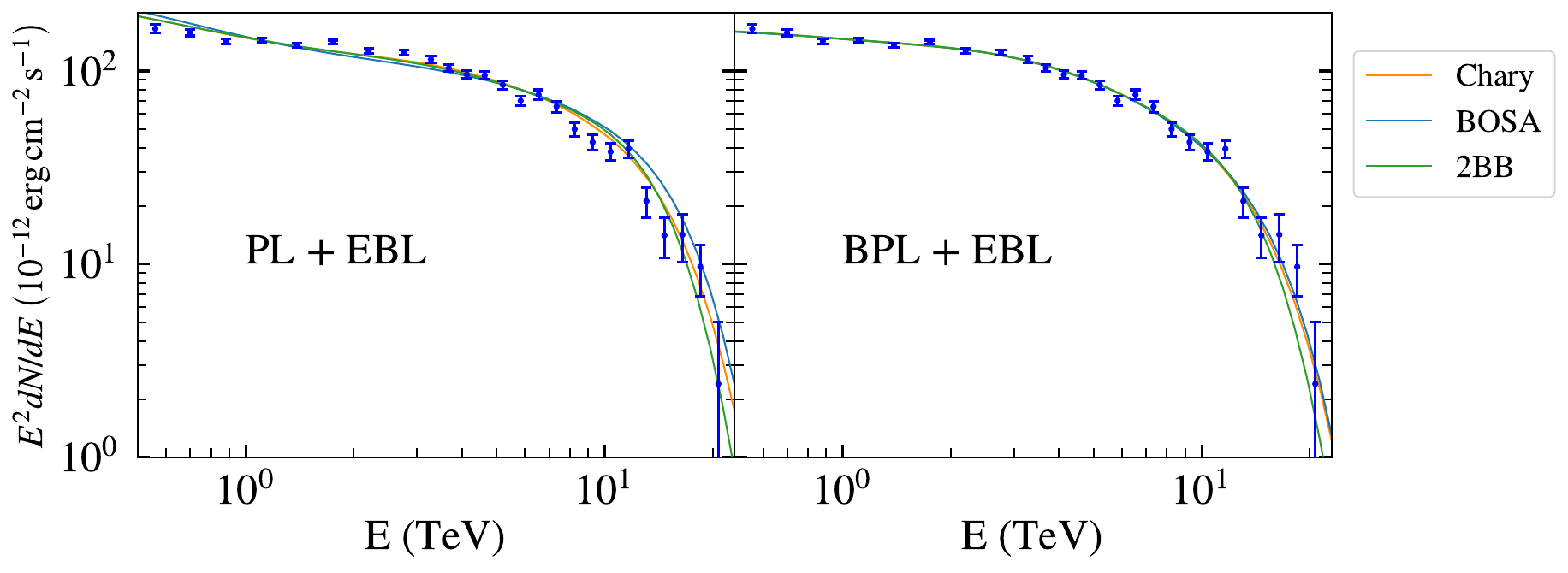}
    \caption{Flare from Markarian~501 in 1997, observed with HEGRA. Data points are taken from the original \citep{1999A&A_FlareInitial} and reanalysis \citep{2001A&AFlareReanalysis}.
    We also show fits for two spectral shapes listed in Table~\ref{tab:spectral_shapes} and the three EBL models calculated in Section~\ref{sect:fitToData}.}
    \label{fig:hegra_spectrum}
\end{figure*}

The spectral shape of the Mkn~501 flare from 1997 is shown in Fig.~\ref{fig:hegra_spectrum}. 
We take the data points from the original analysis \citep{1999A&A_FlareInitial} for energies up to $E = 3\,$TeV, and data from a reanalysis \citep{2001A&AFlareReanalysis} for higher energies. We fit these spectra with the least squares method from \textsc{iminuit}, using different spectral models for the intrinsic spectrum.
The spectral shapes are listed in Table~\ref{tab:spectral_shapes}, and the best-fit parameters in Appendix~\ref{appendix:bestFitsMk501}. 
We test a power law (PL); a smooth broken power law (BPL); a power law with an exponential cutoff (PLE); and a log parabola (LP).
These spectral shapes are then multiplied by the attenuation factor $\exp{(-\tau)}$.
We calculate the $\tau$ of our EBL models, computed in Section~\ref{sect:fitToData}, through numerical integration of Eq.~\eqref{eq:tau} using the python library \textsc{ebltable} \citep{manuel_meyer_2022_ebltable}.

In the expressions of Table~\ref{tab:spectral_shapes}, $\phi_0$ is the normalization of the spectrum, $\Gamma_i$ are the spectral indices, and $E_{0}$ is the pivot energy.
In the BPL $\times$ EBL model, $E_\mathrm{break}$ is the energy where the spectral index changes from $\Gamma_1$ to $\Gamma_2$, and $f$ is the factor that determines how hard this change in index is. We fix $f=2$ in our fits, as allowing it to vary results in overfitting.
In the PLE $\times$ EBL model, $E_\mathrm{cut}$ represents the energy at which the exponential cutoff starts to be relevant. We fix $E_{0}=1\,$TeV to avoid overfitting in the PL, BPL and PLE models. 
Furthermore, $\beta$ is the curvature of the LP model.
The resulting fits, incorporating these three models and the PL and BPL intrinsic spectra, are shown in Fig.~\ref{fig:hegra_spectrum}.

\begin{table}[htb]
    \centering
    \begin{tabular}{rl}
    \midrule
    \midrule
       Model name  &  Spectral shape \\
    \midrule
        PL $\times$ EBL & $\phi (E) = \phi_0 e^{-\tau}  \left(\frac{E}{E_0}\right)^{-\Gamma}$ \\
        BPL $\times$ EBL  & $\phi (E) = \phi_0 e^{-\tau}  \left(\frac{E}{E_0}\right)^{-\Gamma_1}\left[1 + \left(\frac{E}{E_\mathrm{break}}\right)^{f}\right]^{(\Gamma_1-\Gamma_2)/f}$\\
        PLE $\times$ EBL  & $\phi (E) = \phi_0 e^{-\tau}  \left(\frac{E}{E_0}\right) ^{-\Gamma}  e^{-(E / E_\mathrm{cut})}$\\
        LP $\times$ EBL  & $\phi (E) = \phi_0 e^{-\tau}  \left(\frac{E}{E_0}\right) ^ {- \Gamma - \beta  \ln(E / E_0)}$ \\
    \midrule
    \end{tabular}
    \caption{Analytical expressions for the intrinsic spectrum of Mkn~501.}
    \label{tab:spectral_shapes}
\end{table}

The obtained best-fit spectra are then used as the input for a simulation of LHAASO observations.  
We use the Science Book of LHAASO \citep{LHAASOsciencebook} to characterize the detector. 
A flare in the energy range between a few hundred GeV and 20\,TeV would be observed with the water Cherenkov detector array (WCDA), which is sensitive to energies $100\,\mathrm{GeV}$-$30\,\mathrm{TeV}$.
The number of expected counts observed with WCDA is given by
\begin{equation} \label{eq:numberCounts}
\begin{split}
    \mu(\Delta &E_\mathrm{obs}) = \Delta T \int_{\Delta E_\mathrm{obs}} d E_\mathrm{obs} \\
    &\times \int_0^\infty R\left(E_\mathrm{obs}, E_\mathrm{true}\right) A_\mathrm{eff}\left(E_\mathrm{true}\right) \phi \left(E_\mathrm{true}\right) dE_\mathrm{true},
    \end{split}
\end{equation}
where the true energies $E_\mathrm{true}$ are convolved with the energy resolution $R\left(E_\mathrm{obs}, E_\mathrm{true}\right)$ and the effective area $A_\mathrm{eff}\left(E_\mathrm{true}\right)$, and afterwards integrated over the observed energy bins $\Delta E_\mathrm{obs}$. Finally, the result is multiplied by the observation time $\Delta T$.
We choose an effective area that corresponds to the zenith angle interval $0^\circ \leq\theta \leq 15^\circ$.
For the energy resolution of the detector we simply assume a Gaussian function,
\begin{equation}
    R\left(E_\mathrm{obs}, E_\mathrm{true}\right)  = e^{-\frac{1}{2}\frac{\left(E_\mathrm{obs} - E_\mathrm{true}\right)^2}{\sigma_\mathrm{E}^2}},
\end{equation}
and we take an energy resolution $\sigma_\mathrm{E}=0.2E_\mathrm{true}$.
We choose the energy bins so that their number is similar to the number of original HEGRA data points, but this is an arbitrary decision, as we take the energy resolution in the forward folding into account. 
Lastly, we base the observation time on the original flare. The original flare was active from March to October 1997 and was observed for 110\,h with HEGRA \citep{1999A&A_FlareInitial}.
However, since the WCDA is a water Cherenkov detector, it can detect gamma rays from the whole observable sky during the entire day. 
Mkn~501 is at $\theta<30^\circ$ for 4.5\,h a day from the LHAASO coordinates.\footnote{Calculations performed with the help of \url{https://airmass.org/}} Assuming a flare that lasts half a year, we take an approximation of $\Delta T=820\,$h. From this calculation we obtain the expected number of counts as a function of energy, for each spectral shape and EBL model.

To summarize, we assume a) an intrinsic spectral shape based on different spectral models fitted to HEGRA observations, b) that the assumed spectral shape stays constant over the assumed observation time and c) a fixed detector response is similar to that of the LHAASO WCDA detector. 
Furthermore, we do not include a background modeling in our simulations.

In the following, we assume that the true spectral shape of Mkn~501 is a (B)PL $\times$ EBL, and the true EBL is the \textit{Chary} model.
We then evaluate the extent to which different EBL models can be distinguished. 
Subsequently, we present results obtained when adopting alternative true EBL models and different intrinsic spectral shapes for Mkn~501.
We base our calculations on the log-likelihood for Poisson distributed data
\begin{equation}
    -2 \log \mathcal{L}(\mu_i, N_i) = \sum_i-\mu_i + N_i\log\left(\mu_i\right),
\end{equation}
where $\mu_i$ are the expected number of counts and $N_i$ are the number of counts observed in the experiment.
First, we calculate the Asimov data sets \citep{2011EPJC...71.1554C}.
For this data set, we set the observed counts equal to the expected counts $N_i=\mu^\mathrm{As}_i$, where $\mu^\mathrm{As}_i$ is defined as the expected number of counts calculated from injecting the HEGRA spectrum with its best-fit parameters (listed in Appendix~\ref{appendix:bestFitsMk501}) and the true EBL model, using Eq.~\eqref{eq:numberCounts}.

\begin{figure*}[htb]
    \centering
    \includegraphics[width=0.8\linewidth]{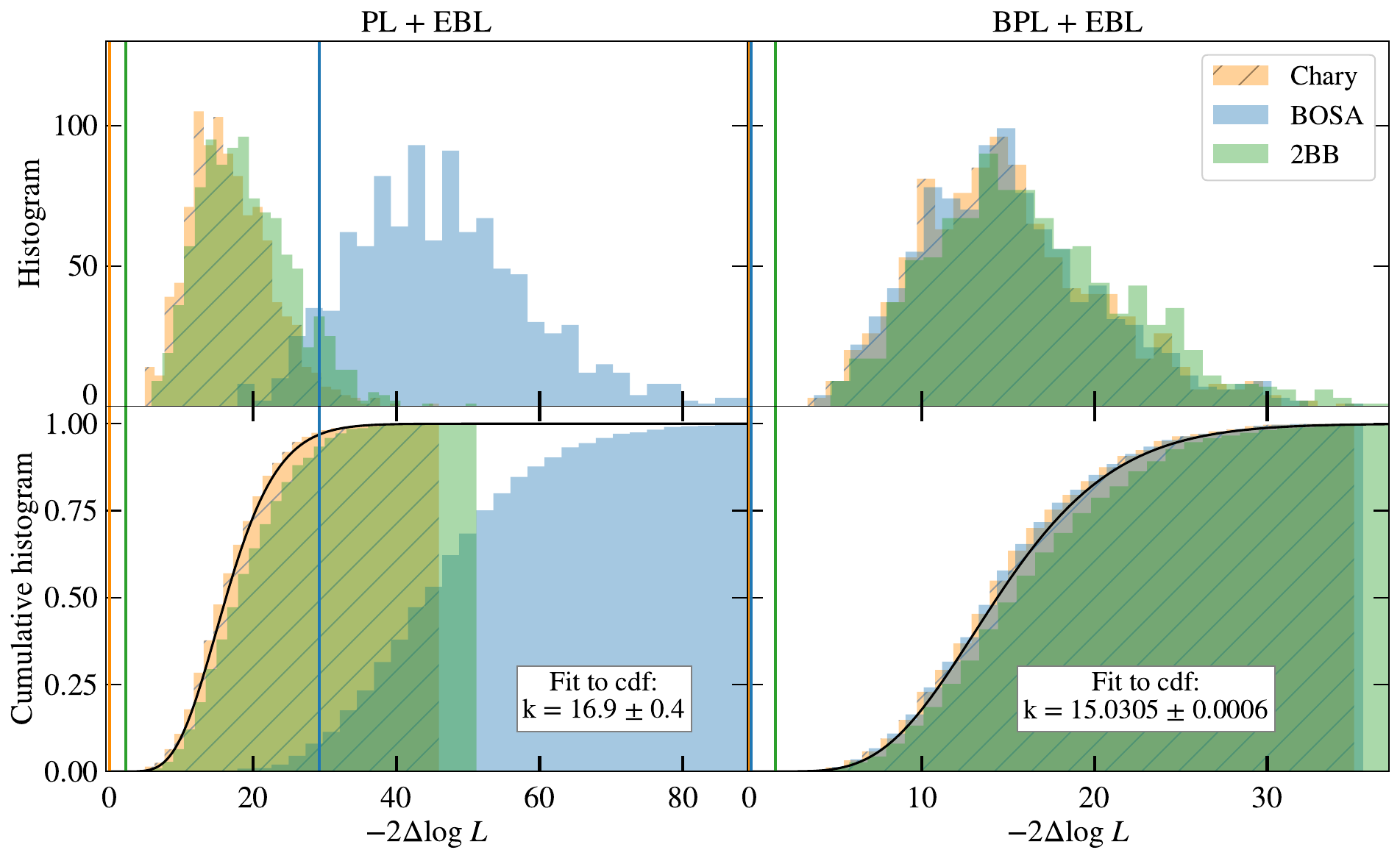}
    \caption{Likelihood ratios of 1000 pseudo-experiments fitting Poissonian draws based on the HEGRA spectrum from the Mkn~501 flare from 1997 with different EBL models and intrinsic spectra. The true assumed EBL model is \textit{Chary}. Vertical lines are the likelihood ratio of the Asimov datasets. \textit{Top:} Histograms of the Poisson draws for the PL$\times$EBL spectral model (\textit{left}) and BPL$\times$EBL spectral model (\textit{right}). \textit{Bottom:} Cumulative distributions. The black lines in the cumulative histograms are the fitted $\chi^2$ distributions of the true EBL histograms, and the best-fit value for the degrees of freedom $k$ is provided as well. Details about the analysis are given in Section~\ref{sect:UseCaseDustProperties}.}
    \label{fig:lhaaso_hist}
\end{figure*}

Now, we investigate the impact of fitting the Asimov dataset with potentially incorrect EBL models. 
We perform fits maximizing the log-likelihood of $-2 \log \mathcal{L}(\mu_i, \mu^\mathrm{As, T}_i)$, where $\mu^\mathrm{As, T}_i$ is the Asimov dataset for the true EBL (fixed values) and $\mu_i$ represent the number of counts that we fit over. The values of $\mu_i$ are calculated from the (B)PL $\times$ EBL spectral shape with an EBL model. The free parameters are the spectral parameters of the intrinsic spectrum which we fit for each of the EBL models we study.
Then, we calculate the ratio
\begin{equation} \label{eq:AsimovRatios}
    -2 \Delta \log \mathcal{L} =  -2 (\log \mathcal{L}(\mu_i, \mu^\mathrm{As, T}_i)+ \log \mathcal{L}(\mu^\mathrm{As, T}_i, \mu^\mathrm{As, T}_i)),
\end{equation}
which represents how well an alternative EBL model explains the Asimov dataset relative to the true EBL model. These ratio values are shown as vertical lines in Fig.~\ref{fig:lhaaso_hist}. We note that the log-likelihood ratio is null when fitting with the true model.

The next step is to perform 1000 Poisson-distributed pseudo-experiments based on $\mu^\mathrm{As, T}_i$.
We denote each of these draws $N^\mathrm{Poiss}_i$.
We repeat the analysis procedure exchanging $\mu^\mathrm{As, T}_i$ with $N^\mathrm{Poiss}_i$.
We perform a fitting maximizing the log-likelihood $-2 \log \mathcal{L}(\mu_i, N^\mathrm{Poiss}_i)$, where $\mu_i$ is the expected number of counts with free spectral parameters, for each of the EBL models. Then, the log-likelihood ratios are calculated as
\begin{equation}
\begin{split}
        -2  \Delta \log \mathcal{L} =  -2 (\log \mathcal{L}(\mu_i, N^\mathrm{Poiss}_i)+ \log \mathcal{L}(N^\mathrm{Poiss}_i, N^\mathrm{Poiss}_i)).
\end{split}
\end{equation}
We present the likelihoods ratio distributions from 1000 pseudo-experiments for our three EBL models and the two spectral shapes in Fig.~\ref{fig:lhaaso_hist}. 
Using the PL $\times$ EBL spectral shape, the distribution of the BOSA ratios does not overlap with the other histograms and exhibits higher values than the other EBL models. This indicates that the fits with the BOSA model are worse assuming \textit{Chary} as the true EBL.
For the BPL $\times$ EBL spectral model, the three models significantly overlap, so we cannot differentiate between them when the spectral model incorporates curvature.

We assume that the ratios follow a $\chi^2$ distribution, and therefore their cumulative functions are given by
\begin{equation} \label{eq:chi2_cumulative}
    \chi^2_{cdf} (-2  \Delta \log L, k) = \frac{\gamma\left(\frac{k}{2}, \frac{-2  \Delta \log L}{2}\right)}{\Gamma\left(\frac{k}{2}\right)},
\end{equation}
where $\gamma\left(\frac{k}{2}, \frac{x}{2}\right)$ is the lower incomplete gamma function, $\Gamma\left(\frac{k}{2}\right)$ is the regularized gamma function, and $k$ denotes the number of degrees of freedom.
We perform a maximum likelihood fit to the cumulative distribution of the true EBL model to the expression in Eq.~\eqref{eq:chi2_cumulative}. We maximize the p-value $p = 1-\chi^2_{cdf}$ treating $k$ as a free parameter, using \textsc{iminuit}. The fit results are shown as black curves on the bottom row of Fig.~\ref{fig:lhaaso_hist}, and the best-fit values of $k$ are also given in the respective panels.
Afterwards, we calculate the p-values of the Asimov ratios (vertical lines, from Eq.~\eqref{eq:AsimovRatios}) against the $\chi^2_{cdf}$ of the true EBL model.
We adopt a threshold $p = 0.05$ to establish a 95\% confidence level. If the calculated p-value falls below this threshold, we reject the null hypothesis that the data can be adequately fitted using an EBL different from the true model, i.e., that we can distinguish the EBL models.

We present the p-values for the PL $\times$ EBL spectral shape in Table~\ref{tab:pl_ebl_pvalues}. The results of using \textit{Chary} as the true EBL model are listed in the second column.
We do not list the p-values for the BPL $\times$ EBL model because the complexity of the model renders all the p-values $p_\mathrm{BPL+EBL} \leq  3 \times 10^{-4}$, and therefore we are unable to distinguish between the models with this analysis. 
Regarding the remaining spectral shapes listed in Table~\ref{tab:spectral_shapes}, their histograms completely overlap for all three EBL models. Therefore, we find all $p_{\mathrm{LP} \times \mathrm{EBL}} \leq 2 \times 10^{-4}$, and all $p_{\mathrm{PLE} \times \mathrm{EBL}} \leq 7 \times 10^{-7}$.
The curvature in these spectral shapes allows the spectral parameters to compensate for the differing optical depths, preventing any distinction between the EBL models using the likelihood ratios.

\begin{figure*}[htb]
    \centering
    \includegraphics[width=0.99\linewidth]{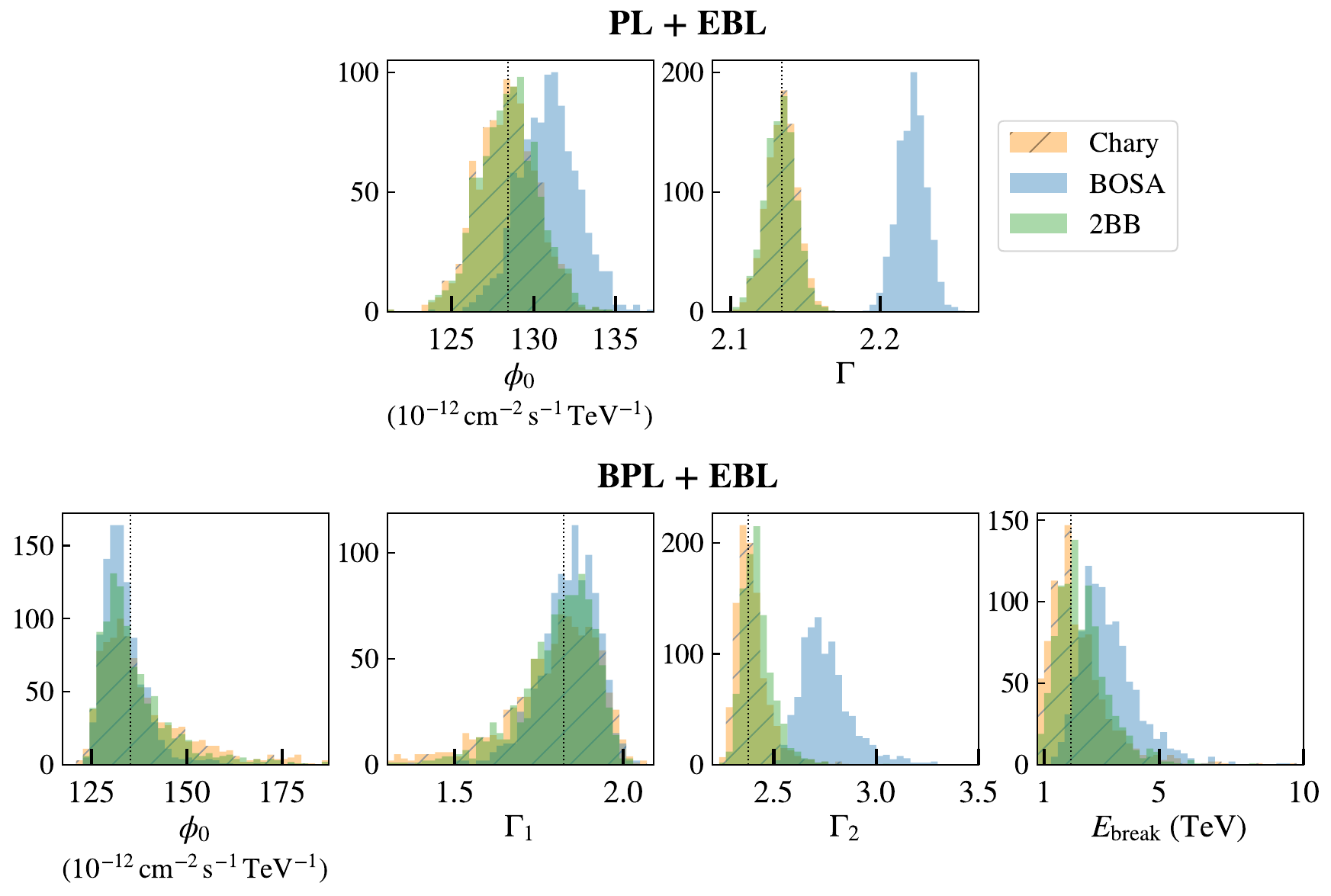}
    \caption{Best-fit parameters for the 1000 Poissonian pseudo-experiments assuming \textit{Chary} is the true EBL model, for the fits performed with the three EBL models from Fig.~\ref{fig:ebl_fits}. Dotted vertical lines are the best-fit parameters of the Asimov dataset. Parameters are given for the two spectral shapes studied in this work, listed in Table~\ref{tab:spectral_shapes}.}
    \label{fig:lhaaso_params}
\end{figure*}

In Fig.~\ref{fig:lhaaso_params}, we show the histograms of the best-fit parameters for the 1000 pseudo-experiments assuming \textit{Chary} as the true model. The histograms of the \textit{Chary} and 2BB models overlap for the PL $\times$ EBL spectral shape, but not for BOSA. For the BPL $\times$ EBL, some parameters allow for discrimination between models: $\phi_0$ and $\Gamma_1$ discriminate \textit{Chary} from the other models, while $\Gamma_2$ can be used to discriminate BOSA model.

\begin{table}
    \centering
    \begin{tabular}{c|ccc}
    \midrule
    \midrule
    \backslashbox{True}{Assumed}
         & BOSA & \textit{Chary} & 2BB \\
    \midrule
        BOSA  & -        & 0.977445 & 0.972446 \\
        \textit{Chary} & 0.969198 & -        & 0.000008 \\
        2BB   & 0.954735 & 0.000008 & - \\
    \midrule
    \end{tabular}
    \caption{P-values for the PL$\times$EBL spectral shape. 
    P-values below the threshold $p = 0.05$ reject the null hypothesis that the pseudo-experiments can be adequately fitted using an EBL different from the true model.
    }
    \label{tab:pl_ebl_pvalues}
\end{table}

\section{Conclusions and outlook} \label{sect:conclusions}

In this work we present \textsc{niebla}, an open source \textsc{Python} library designed to calculate the EBL. We follow a phenomenological approach, characterizing the sources that contribute to the EBL intensity and evolving their luminosities through cosmic time. The code already provides various parametrizations and templates to simulate the typical sources contributing to the EBL intensity, and also can incorporate custom inputs.

We present the results of modeling and fitting the EBL to observational data using three different dust reemission techniques. Two of them are based on synthetic templates, and the last one on black-body reemission. Using either the \textit{Chary} templates or the black bodies yields the best fit, indicated by their reduced $\chi^2$ values. The fitting results are primarily driven by the emissivity data in the optical regime.

Our EBL models can be compared to established benchmark EBL models from the literature, such as those presented by \citet{finke_modeling_2022}, \citet{saldana-lopez_observational_2021}, and \citet{2017A&A...603A..34F}.
In the optical regime, there is consistency in spectral shapes among all models.
Differences become more pronounced in the infrared regime, primarily due to the different dust reemission methodologies and the reduced number of measurements in this regime. A more detailed comparison is given in Appendix~\ref{appendix:EBL_figure}.

We also present a use case for the code by analyzing how a VHE observation would help us distinguish between different EBL models. We simulate an observation, similar to the 1997 flare from Mkn~501 with a LHAASO-like detector. 
Our analysis shows that this observation would allow us to distinguish between BOSA and the other EBL models if the intrinsic spectrum of the flare was a simple PL.
However, we cannot differentiate between the \textit{Chary} and 2BB models.
Furthermore, when the intrinsic spectral model includes curvature, 
we are unable to differentiate different dust prescriptions in the EBL modeling. 
The potential of this methodology would increase by analyzing multiple flares from various sources, which would enable us to differentiate EBL models with higher precision.
It should also be noted that these findings are based on several assumptions, such as a constant spectral shape over the observation time and a negligible contribution of background counts in the detector.
Nevertheless, 
our EBL modeling code provides the necessary tools for future studies of the EBL using VHE observations.
Specifically, probing the EBL via its opacity effects in VHE observations constitutes one of the cosmology case studies for the CTAO \citep{2021JCAP...02..048A}. Currently, this methodology relies on fitting the overall normalization of the EBL or simplifying its spectral shape to a parametric expression. \textsc{Niebla} enables us to study the underlying physical parameters of the EBL.

\section*{Acknowledgments}
The authors would like to thank Alberto Domínguez for his helpful comments on the manuscript, Dieter Horns for his thoughtful comments and discussions, and Atreya Acharyya for testing the \textsc{niebla} tutorials.
The authors acknowledge support from the European Research Council (ERC) under the European Union’s Horizon2020 research and innovation program Grant Agreement No. 948689 (AxionDM) and from the Deutsche Forschungsgemeinschaft (DFG, German Research Foundation) under Germany’s Excellence Strategy – EXC 2121 “Quantum Universe”– 390833306. This article is based upon work from COST Action COSMIC WISPers CA21106, supported by COST (European Cooperation in Science and Technology).
This research has made use of the Astrophysics Data System, funded by NASA under Cooperative Agreement 80NSSC21M00561. It has also made use of \textsc{python}, along with community-developed or maintained software packages, including \textsc{matplotlib} \cite{2007CSE.....9...90H}, \textsc{numpy} \cite{harris2020array}, \textsc{scipy} \cite{2020NatMe..17..261V}, \textsc{astropy} \cite{2013A&A...558A..33A,2018AJ....156..123A,2022ApJ...935..167A} and \textsc{iminuit} \cite{iminuit}.

\bibliographystyle{aa}
\bibliography{library}
\appendix 

\section{Parametrizations already listed in the EBL code} \label{appendix:parametrizationsInCode}
We present a list of all the parametrizations present in the EBL code in Table~\ref{tab:parametrizationsInCode}. The possibility to input custom parameters and functions is also available for all the expressions that we list here.

\begin{table*}[htb]
\centering
\resizebox{\textwidth}{!}{
\begin{tabular}{lll}
\toprule \toprule
Reference & Name in the code & Expression \\ 

\toprule \toprule
\multicolumn{3}{c}{$\rho_{\star}$} \\ 
\midrule

Eq.~15 from \citet{madau_cosmic_2014} & sfr\_madau14 & $\rho_{\star}(z) = x_0 \frac{(1 + z)^{x_1}}{1 + [(1+z)/x_2]^{x_3}}$ \\
Eq.~53 from \citet{haardt_radiative_2012} & sfr\_cuba & $\rho_{\star}(z) = \frac{x_0 + x_1 (z / x_2) ^ {x_3}}{1 + (z / x_4) ^ {x_5}}$ \\
Eq.~15 from \citet{finke_modeling_2022} & sfr\_finke22a & See original paper \\
Constant (or none) & sfr\_constant & $\rho_{\star}(z) = x_0$ \\

\midrule \midrule   
\multicolumn{3}{c}{$Z(z)$} \\
\midrule

Eq.~4 from \citet{2022Tanikawa} & metall\_tanikawa22 & $Z (z) = Z_\odot \times 10^{a_0 - a_1 \times z ^ {a_2}}$  \\
Constant  & constant & $Z(z) = a_0$ \\

\midrule \midrule
\multicolumn{3}{c}{Dust absorption} \\
\midrule
Wavelength dependency \\
\cmidrule(lr){1-1}
Eq.~6 from \citet{kneiske_implications_2002} & kneiske2002 & $f_\mathrm{esc, dust} (\lambda) = 10^{-0.4\,A_{\lambda}}$    with $A_{\lambda} = 0.68 \cdot E(B-V) \cdot R \cdot (\lambda^{-1}-0.35)$ \\
Eq.~15 from \citet{razzaque_stellar_2009} & razzaque2009 & See original paper\\
Eq.~13 from \citet{finke_modeling_2022} & dust\_att\_finke & See original paper \\
\midrule
Redshift dependency \\
\cmidrule(lr){1-1}
Appendix from  & fermi2018 & $f_\mathrm{esc, dust}(z) = 10^{-0.4 A(z)}$    with $A(z) = m_d \frac{(1 + z)^{n_d}}{1 + [(1+z)/p_d]^{q_d}}$\\
\citet{the_fermi-lat_collaboration_gamma-ray_2018}\\
\midrule
Combined dependency \\
\cmidrule(lr){1-1}
razzaque2009 $\times$ fermi2018 & comb\_model\_1 \\
dust\_att\_finke $\times$ fermi2018 & finke2022 \\
\midrule
\end{tabular}
}
\caption{List of parametrizations already included in the EBL modeling code.
}
\label{tab:parametrizationsInCode}
\end{table*}

\section{Best-fit parameters and correlation matrices} \label{appendix:correlation_matrices}

In Table~\ref{tab:bestfits_nosystematics} we present the best-fit parameters of the EBL models obtained assuming no systematic uncertainties.

For each of the three EBL models, and for both fitting scenarios, with and without added systematics, we provide the corresponding correlation matrices. Each matrix is derived from the covariance matrix evaluated at the minimum of the $\chi^2$ function. The diagonal elements of the covariance matrices correspond to the parameter variances and have already been listed as uncertainty estimates in Table~\ref{tab:bestfits} and Table~\ref{tab:bestfits_nosystematics}. Here, we list the correlation matrices, defined as  
$\mathrm{corr}_\mathrm{i, j} = \mathrm{cov}_\mathrm{i, j}/(\sigma_\mathrm{i}\sigma_\mathrm{j})$. 
For the case without added systematics, the correlation matrices are given for the \textit{Chary} model in Table~\ref{tab:correlation_chary_nosys}, BOSA in Table~\ref{tab:correlation_bosa_nosys}, and 2BB in Table~\ref{tab:correlation_2bb_nosys}. For the case with 14~\% added systematics, the correlation matrices are given for the \textit{Chary} model in Table~\ref{tab:correlation_chary_14sys}, BOSA in Table~\ref{tab:correlation_bosa_14sys}, and 2BB in Table~\ref{tab:correlation_2bb_14sys}.

\begin{table*}[htb]
    \centering
    \begin{tabular}{cccc} \midrule\midrule
        Parameter & BOSA & Chary & 2BB \\ \midrule
        \multicolumn{4}{c}{$\rho_{\star}$} \\ \midrule
        $x_0$ $\left( \times 10^{-2} \,\mathrm{M}_{\odot} / \mathrm{yr} / \mathrm{Mpc}^{3}\right)$ & $1.5 \pm 0.5$ & $1.29\pm 0.03$ & $1.24\pm 0.03$ \\
        $x_1$ & $2.25\pm 0.05$  & $2.43\pm 0.05$ & $2.44\pm 0.04$\\
        $x_2$ & $3.35\pm 0.04$  & $3.31\pm 0.04$ & $3.33\pm 0.04$\\
        $x_3$ &  $6.56\pm0.07$  & $6.70\pm 0.07$ & $6.73\pm 0.07$\\ \midrule
        \multicolumn{4}{c}{\textit{Z}} \\ \midrule
        $a_0$ & $-0.15\pm0.12$  & $-0.28\pm 0.05$ & $-0.54\pm 0.04$\\
        $a_1$ &  $0.21\pm0.09$  & $0.20\pm 0.05$ & $0.025\pm 0.018$\\
        $a_2$ & $0.82\pm0.14$  & $0.66\pm 0.14$ & $1.6\pm 0.4$\\ \midrule
        \multicolumn{4}{c}{Dust reemission} \\ \midrule
        $f_\mathrm{TIR}$ $\left(\mathrm{M}_\odot^{-1}\right)$ (dex) & - &  $9.71\pm 0.03$ & - \\
        $\lambda_\mathrm{TIR, cut}$ (µm) & - &  $5.36\pm 0.02$ & - \\
        $f_1$ & - & - & $0.167\pm 0.007$ \\
        $T_2$ (K) & - & - & $62.9\pm 0.9$ \\
        \midrule
    \end{tabular}
    \caption{
    Best-fit parameters for the EBL models without systematic uncertainties, detailed in Section~\ref{sect:fitToData}.
    }
    \label{tab:bestfits_nosystematics}
\end{table*}

\begin{table*}[]
    \centering
    \begin{tabular}{cccccccccc}
\toprule
\hline
& $x_0$ & $x_1$ & $x_2$ & $x_3$ & $a_0$ & $a_1$ & $a_2$ & $f_\mathrm{TIR}$ & $\lambda_\mathrm{TIR, cut}$ \\
\hline
$x_0$ & 1 &  &  &  &  &  &  &  & \\
$x_1$ & -0.89 & 1 &  &  &  &  &  &  & \\
$x_2$ & 0.52 & -0.80 & 1 &  &  &  &  &  &  \\
$x_3$ & -0.19 & -0.02 & 0.57 & 1 &  &  &  &  &  \\
$a_0$ & 0.43 & -0.35 & 0.15 & -0.15 & 1 &  &  &  & \\
$a_1$ & 0.11 & -0.18 & 0.11 & -0.10 & 0.84 & 1 &  &  &  \\
$a_2$ & 0.12 & -0.01 & -0.03 & 0.09 & -0.55 & -0.83 & 1 &  & \\
$f_\mathrm{TIR}$  & 0.07 & -0.04 & 0.01 & -0.02 & 0.04 & 0.02 & -0.00 & 1 &  \\
$\lambda_\mathrm{TIR, cut}$ & -0.01 & 0.01 & -0.00 & 0.01 & 0.04 & 0.02 & -0.00 & 0.05 & 1 \\
\hline
\end{tabular}
    \caption{Correlation coefficients between the best-fit parameters for the \textit{Chary} model, for the case without systematic uncertainty in the data points. Since the matrix is symmetric, the upper trigonal part is not listed.}
    \label{tab:correlation_chary_nosys}
\end{table*}

\begin{table*}[]
    \centering
    \begin{tabular}{cccccccc}
\toprule
\hline
& $x_0$ & $x_1$ & $x_2$ & $x_3$ & $a_0$ & $a_1$ & $a_2$ \\
\hline
$x_0$ & 1 &  &  &  &  &  &  \\
$x_1$  & -0.90 & 1 &  &  &  &  &  \\
$x_2$ & 0.42 & -0.71 & 1 &  &  &  &  \\
$x_3$ & -0.30 & 0.10 & 0.55 & 1 &  &  &  \\
$a_0$ & 0.84 & -0.65 & 0.17 & -0.28 & 1 &  & \\
$a_1$ & 0.77 & -0.63 & 0.19 & -0.28 & 0.97 & 1 &  \\
$a_2$ & -0.57 & 0.48 & -0.15 & 0.28 & -0.78 & -0.87 & 1\\
\hline
\end{tabular}
    \caption{Correlation coefficients between the best-fit parameters for the BOSA model, for the case without systematic uncertainty in the data points. Since the matrix is symmetric, the lower trigonal part is not listed.}
    \label{tab:correlation_bosa_nosys}
\end{table*}

\begin{table*}[]
    \centering
    \begin{tabular}{cccccccccc}
\toprule
\hline
& $x_0$ & $x_1$ & $x_2$ & $x_3$ & $a_0$ & $a_1$ & $a_2$ & $T_2$ & $f_1$ \\
\hline
$x_0$ & 1 &  &  &  &  &  &  &  & \\
$x_1$ & -0.88 & 1 &  &  &  &  &  &  & \\
$x_2$ & 0.48 & -0.78 & 1 &  &  &  &  &  & \\
$x_3$ & -0.17 & -0.04 & 0.59 & 1 &  &  &  &  & \\
$a_0$ & 0.61 & -0.35 & 0.06 & -0.13 & 1 &  &  &  & \\
$a_1$ & 0.28 & -0.20 & -0.00 & -0.21 & 0.64 & 1 &  &  & \\
$a_2$ & -0.11 & 0.09 & 0.05 & 0.28 & -0.38 & -0.90 & 1 &  & \\
$T_2$ & 0.19 & -0.19 & 0.12 & -0.03 & -0.04 & -0.05 & 0.04 & 1 & \\
$f_1$ & -0.07 & 0.02 & 0.01 & 0.00 & -0.01 & 0.03 & -0.03 & -0.27 & 1\\
\hline
\end{tabular}
    \caption{Correlation coefficients between the best-fit parameters for the 2BB model, for the case without systematic uncertainty in the data points. Since the matrix is symmetric, the lower trigonal part is not listed.}
    \label{tab:correlation_2bb_nosys}
\end{table*}

\begin{table*}[]
    \centering
    \begin{tabular}{cccccccccc}
\toprule
\hline
& $x_0$ & $x_1$ & $x_2$ & $x_3$ & $a_0$ & $a_1$ & $a_2$ & $f_\mathrm{TIR}$ & $\lambda_\mathrm{TIR, cut}$ \\
\hline
$x_0$ & 1 &  &  &  &  &  &  &  & \\
$x_1$ & -0.86 & 1 &  &  &  &  &  &  & \\
$x_2$ & 0.56 & -0.87 & 1 &  &  &  &  &  & \\
$x_3$ & -0.33 & 0.07 & 0.36 & 1 &  &  &  &  & \\
$a_0$ & 0.18 & -0.16 & 0.09 & -0.10 & 1 &  &  &  & \\
$a_1$ & 0.07 & -0.14 & 0.11 & -0.07 & 0.89 & 1 &  &  & \\
$a_2$ & 0.02 & 0.04 & -0.04 & 0.08 & -0.70 & -0.80 & 1 &  & \\
$f_\mathrm{TIR}$ & 0.04 & -0.02 & 0.01 & -0.01 & 0.09 & 0.05 & -0.01 & 1 &  \\
$\lambda_\mathrm{TIR, cut}$ & -0.02 & 0.01 & -0.00 & 0.01 & 0.00 & -0.00 & 0.00 & -0.01 & 1 \\
\hline
\end{tabular}
    \caption{Correlation coefficients between the best-fit parameters for the \textit{Chary} model, for the case of 14\% uncertainty in the data points. Since the matrix is symmetric, the upper trigonal part is not listed.}
    \label{tab:correlation_chary_14sys}
\end{table*}

\begin{table*}[]
    \centering
    \begin{tabular}{cccccccc}
\toprule
\hline
& $x_0$ & $x_1$ & $x_2$ & $x_3$ & $a_0$ & $a_1$ & $a_2$ \\
\hline
$x_0$ & 1 &  &  &  &  &  &  \\
$x_1$ & -0.85 & 1 &  &  &  & & \\
$x_2$ & 0.51 & -0.85 & 1 &  &  &  & \\
$x_3$ & -0.30 & 0.04 & 0.40 & 1 &  &  &  \\
$a_0$ & 0.39 & -0.23 & 0.08 & -0.13 & 1 &  &   \\
$a_1$ & 0.18 & -0.18 & 0.11 & -0.09 & 0.79 & 1 &   \\
$a_2$ & 0.09 & 0.00 & -0.04 & 0.04 & -0.28 & -0.66 & 1  \\
\hline
\end{tabular}
    \caption{Correlation coefficients between the best-fit parameters for the BOSA model, for the case of 14\% uncertainty in the data points. Since the matrix is symmetric, the lower trigonal part is not listed.}
    \label{tab:correlation_bosa_14sys}
\end{table*}

\begin{table*}[]
    \centering
    \begin{tabular}{cccccccccc}
\toprule
\hline
& $x_0$ & $x_1$ & $x_2$ & $x_3$ & $a_0$ & $a_1$ & $a_2$ & $T_2$ & $f_1$ \\
\hline
$x_0$ & 1 &  &  &  &  &  &  &  & \\
$x_1$ & -0.82 & 1 &  &  &  &  &  & \\
$x_2$ & 0.46 & -0.85 & 1 &  &  &  &  &  & \\
$x_3$ & -0.31 & 0.07 & 0.37 & 1 &  &  &  &  &  \\
$a_0$ & 0.39 & -0.19 & 0.02 & -0.15 & 1 &  &  &  & \\
$a_1$ & 0.07 & -0.17 & 0.13 & -0.13 & 0.65 & 1 &  &  &  \\
$a_2$ & 0.10 & 0.09 & -0.13 & 0.11 & -0.31 & -0.87 & 1 &  & \\
$T_2$ & 0.15 & -0.13 & 0.08 & -0.05 & -0.07 & -0.03 & 0.01 & 1 & \\
$f_1$ & -0.05 & -0.02 & 0.04 & 0.01 & 0.03 & 0.02 & -0.01 & -0.23 & 1 \\
\hline
\end{tabular}
    \caption{Correlation coefficients between the best-fit parameters for the 2BB model, for the case of 14\% uncertainty in the data points. Since the matrix is symmetric, the lower trigonal part is not listed.}
    \label{tab:correlation_2bb_14sys}
\end{table*}

\section{EBL evolution with redshift} \label{appendix:EBL_figure}

\begin{figure*}[htb]
    \centering
    \includegraphics[width=0.99\linewidth]{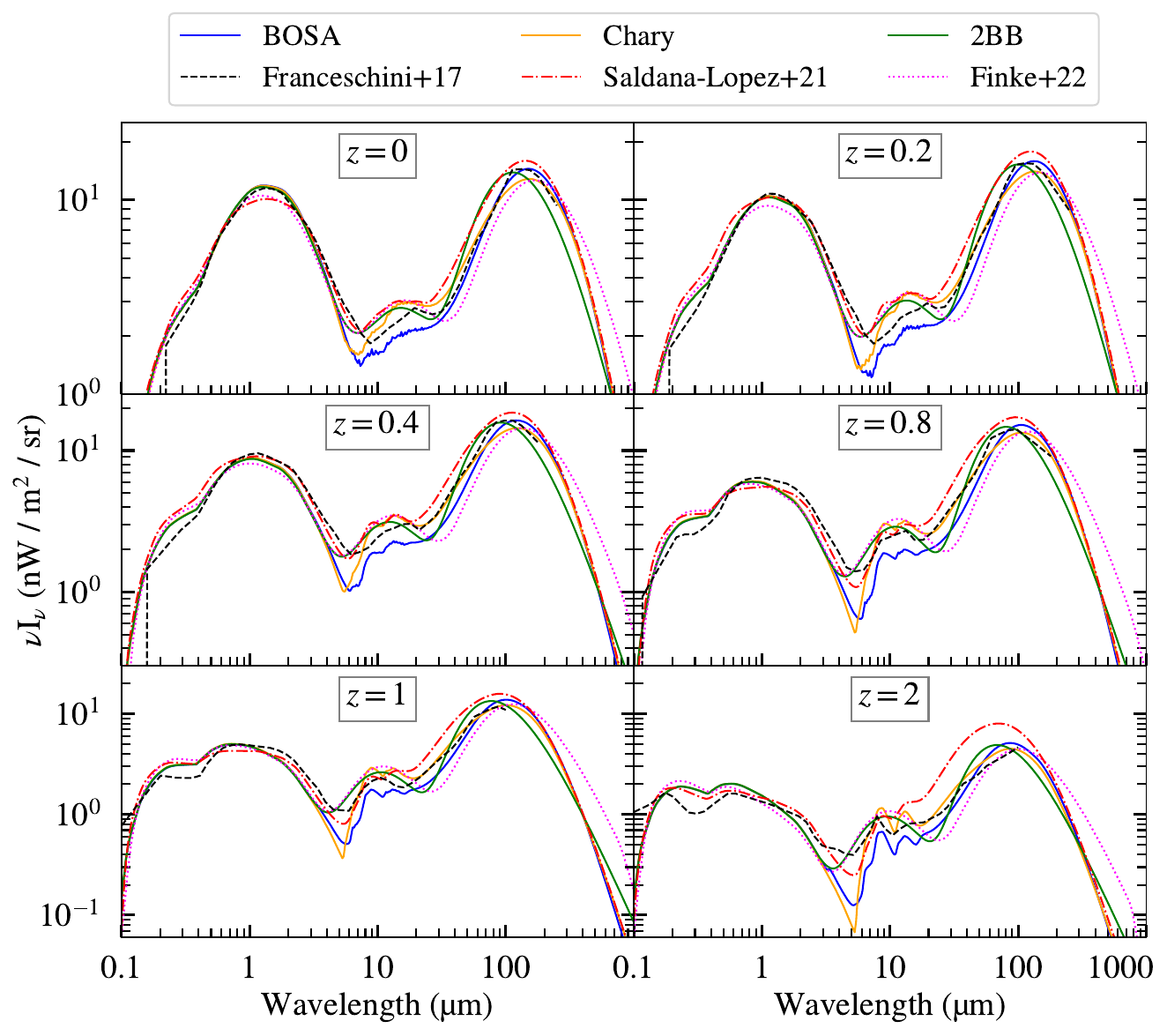}
    \caption{EBL spectral shape for different redshifts. We provide the evolution for our three EBL models (straight lines), as well as benchmark models like \citet{2017A&A...603A..34F} (black dashed lines), \citet{saldana-lopez_observational_2021} (red dash-dotted lines) and \citet{finke_modeling_2022} (fuchsia dotted lines). \citet{2017A&A...603A..34F} incorporate the intensity of the cosmic microwave background at larger wavelengths, which we do not show to facilitate comparison between the models.}
    \label{fig:ebl_redshifts}
\end{figure*}

In Fig.~\ref{fig:ebl_redshifts} we provide the evolution of our EBL models as a function of redshift, and the comparison with some benchmark EBL models. We use a phenomenological approach, which is the same methodology that \citet{finke_modeling_2022} use. We also share the dust absorption methodology, so the spectral shapes are very similar in the optical regime. Moreover, our 2BB model shares their methodology of dust reemission in the MIR, so these two are similar in this regime.
\citet{saldana-lopez_observational_2021} derive the EBL from observed spectral energy densities of galaxies without fits to the actual EBL observational data itself.
This model predicts slightly higher intensity in the CIB.
The model from \citet{2017A&A...603A..34F} incorporates the intensity of the cosmic microwave background, which is the dominant contribution at larger wavelengths. To facilitate comparison between models, we do not show this model at large wavelengths.

\section{Changing the cosmological model} \label{appendix:cosmology}

By default, \textsc{niebla} employs a standard $\Lambda$CDM cosmology with parameters
$H_0 = 70\,\mathrm{km}\,\mathrm{s}^{-1}\,\mathrm{Mpc}^{-1}$, $\Omega_\mathrm{matter}=0.3$, $\Omega_\Lambda=0.7$, and $\Omega_\mathrm{baryons}=0.0453$.
However, the user is free to modify these parameters or use another cosmological model entirely.
Internally, \textsc{niebla} relies on the cosmology classes of \textsc{astropy} \cite{2013A&A...558A..33A,2018AJ....156..123A,2022ApJ...935..167A} and stores the cosmology in a protected variable.
The values of the $\Lambda$CDM parameters can be changed via the \textit{yaml} file or dictionary with the inputs of the code as further detailed in the example notebooks.  
A change of the cosmological model itself is possible as well, as long as the user uses the same object structure as the one already provided. 
We have tested that the other \textsc{astropy} cosmology objects work, and in addition, the user can provide their own custom class, which should be stored in the
\verb|EBL_model._cosmo| 
variable. 
A custom class, called
\verb|custom_cosmology| 
hereafter, needs to include the following functions, which are necessary for the EBL calculation:
\begin{itemize}
\item[\textbullet]
For the general calculation regarding evolution of stellar spectra:
\begin{itemize}
  \item[$\rightarrow$] \verb|custom_cosmology.lookback_time(z)| 
  - calculation of time as a function of redshift $z$, which can be given as a
  \verb|float| 
  or an
  \verb|array|.
  \item[$\rightarrow$] \verb|custom_cosmology.H(z)| 
  - calculation of the Hubble parameter as a function of redshift $z$, which can be given as a 
  \verb|float|
  or an 
  \verb|array|.
\end{itemize}
\item[\textbullet] 
For the calculation of an ALP contribution:
\begin{itemize}
  \item[$\rightarrow$] \verb|custom_cosmology.Odm(z)| 
  - calculate the density parameter for dark matter  as a function of redshift $z$, which can be given as a
  \verb|float|
  or an 
  \verb|array|.
  Only the case for $z=0$ is needed, but the function must accept an input for compatibility.
  \item[$\rightarrow$] \verb|custom_cosmology.critical_density0| 
  - value of the critical mass density at $z=0$.
\end{itemize}

\item[\textbullet] 
For the IHL calculation, it is recommended to use the \textsc{astropy} class, since the library \textsc{hmf} internally wraps the cosmology object:
\begin{verbatim}
from hmf import MassFunction
[...]
mf = MassFunction(cosmo_model=self._cosmo, 
                  Mmin=m_min, Mmax=m_max)
\end{verbatim}
\end{itemize}

All these parameters must be 
\verb|Quantity| 
objects from the 
\verb|astropy.units| 
module.

Modifying the cosmological model must be made with caution. The data reduction pipelines for several observables, including $\rho_{\star}$ and $\varepsilon_{\nu}$, explicitly assume a specific cosmology. 
Consequently, any fitting procedure or comparison of EBL models that employs a nonstandard cosmology must be interpreted carefully. 
If necessary, it is the responsibility of the user to rescale the data inputs to different cosmologies as done in, e.g., \citet{Dominguez_2019,finke_modeling_2022,Dominguez_2024}.

\section{Best-fit parameters of the Mkn~501 spectrum} \label{appendix:bestFitsMk501}
We present a list with the best-fit parameters of the Mkn~501 fits of the HEGRA observations from the flare from 1997.

\begin{table*}[htb]
\centering
\begin{tabular}{lcccccl}
\toprule \toprule
\multicolumn{7}{c}{PL $\times$ EBL} \\
\toprule
& $\phi_0$ $\left(\times 10^{-12}\, \mathrm{cm}^{-2}\,\mathrm{ s}^{-1}\,\mathrm{ TeV}^{-1}\right)$ & $\Gamma$ &&& $\chi^2/\mathrm{d.o.f.}$  & Fixed parameters:\\
\cmidrule(lr){2-3}\cmidrule(lr){6-6}
BOSA & $ 130 \pm 2 $ & $2.20 \pm 0.01$ &&& $161.7 / 23 = 7.03 $ & $E_0 = 1\,$TeV \\ 
\textit{Chary} & $ 128 \pm 2 $ & $2.13 \pm 0.01$ &&& $ 80.00 / 23 = 3.48 $ \\ 
2BB & $ 128 \pm 2 $ & $ 2.13 \pm 0.01$ &&&  $ 93.07 / 23 = 4.05 $ \\

\toprule \toprule
\multicolumn{7}{c}{BPL $\times$ EBL}  \\
\toprule
& $\phi_0$ $\left(\times 10^{-12}\, \mathrm{cm}^{-2}\,\mathrm{ s}^{-1}\,\mathrm{ TeV}^{-1}\right)$ & $\Gamma_1$ & $\Gamma_2$ & $E_\mathrm{break}$ $\left(\mathrm{TeV}\right)$ & $\chi^2/\mathrm{d.o.f.}$   & Fixed parameters:\\
\cmidrule(lr){2-5}\cmidrule(lr){6-6}
BOSA & $134 \pm 4$ & $1.82 \pm 0.07$ & $2.66 \pm 0.09$ & $2.7 \pm 0.6$ & $25.32 / 21 = 1.21$ & $E_0 = 1\,$TeV\\
\textit{Chary} & $135 \pm 6$ & $1.8 \pm 0.1$ & $2.38 \pm 0.06$ & $1.9 \pm 0.6$ & $26.37 / 21 = 1.26 $ & $f = 2$\\
2BB & $132 \pm 4$ & $1.84 \pm 0.07$ & $2.44 \pm 0.08$ & $2.4 \pm 0.7$ & $28.211 / 21 = 1.34$ \\

\toprule \toprule
\multicolumn{7}{c}{PLE $\times$ EBL}  \\
\toprule
& $\phi_0$ $\left(\times 10^{-12}\, \mathrm{cm}^{-2}\,\mathrm{ s}^{-1}\,\mathrm{ TeV}^{-1}\right)$ & $\Gamma$ & $E_\mathrm{cut}$ $\left(\mathrm{TeV}\right)$ && $\chi^2/\mathrm{d.o.f.}$   & Fixed parameters:\\
\cmidrule(lr){2-4}\cmidrule(lr){6-6}
BOSA  & $ 141 \pm 2 $ & $ 1.90 \pm 0.03 $ & $ 10 \pm 1 $ && $40.56 / 22 = 1.84 $ & $E_0 = 1\,$TeV \\
\textit{Chary} & $ 135 \pm 2 $ & $ 1.96 \pm 0.03 $ & $ 16 \pm 3 $ && $ 38.94 / 22 = 1.77 $\\
2BB & $ 135 \pm 2 $ & $ 1.92 \pm 0.03 $ & $ 14 \pm 2 $ && $ 39.17 / 22 = 1.78 $\\

\toprule \toprule
\multicolumn{7}{c}{LP $\times$ EBL}\\
\toprule
& $\phi_0$ $\left(\times 10^{-12}\, \mathrm{cm}^{-2}\,\mathrm{ s}^{-1}\,\mathrm{ TeV}^{-1}\right)$ & $\Gamma$ & $E_0$ $\left(\mathrm{TeV}\right)$ & $\beta$ & $\chi^2/\mathrm{d.o.f.}$\\
\cmidrule(lr){2-5}\cmidrule(lr){6-6}
BOSA   & $ 88 \pm 10 $ & $ 1.99 \pm 0.03 $ & $ 1.21 \pm 0.06 $ & $ 0.16 \pm 0.02 $ & $29.33 / 21 = 1.40 $\\
\textit{Chary}  & $ 100 \pm 11 $ & $ 2.00 \pm 0.03 $ & $ 1.12 \pm 0.06 $ & $ 0.09 \pm 0.02 $ & $29.56 / 21 = 1.41 $\\
2BB  & $ 101 \pm 11 $ & $ 1.97 \pm 0.03 $ & $ 1.12 \pm 0.06 $ & $ 0.10 \pm 0.02 $ & $31.38 / 21 = 1.49 $\\
\midrule
\end{tabular}
\caption{List of best-fit parameters of the 1997 flare from Mkn~501 observed with HEGRA. We also list the $\chi^2$ values and fixed parameters for each spectral model. The EBL models are our best fit ones, assuming a 14~\% systematic uncertainty in all data points.}
\label{tab:fitsMk501}
\end{table*}

\end{document}